\documentclass[12pt, openright]{book}
\usepackage{uorthesis}
\usepackage{booktabs} 
\usepackage{subcaption}
\usepackage{xcolor}

\usepackage{graphicx}
\usepackage{tikz}
\usepackage{subcaption}
\usepackage{caption}
\usepackage{multirow}

\usepackage[normalem]{ulem}

\usepackage{float}

\begin{document}
\pagenumbering{roman}

%% Simple title page

% \begin{titlepage}
%     \begin{center}
%         \vfill
        
%         \Huge
%         {Chasing Ghost Particles: Searching for High-Energy Neutrinos from Core-Collapse Supernovae}
        
%         \vspace{0.5cm}
%         \Large
%         by\\
    
%         {Your Name}
        
%         \vfill

%         \large
%         A thesis submitted in partial fulfillment \\
%         of the requirements for the degree of\\
%         Bachelor of Science\\
%         Physics and Astronomy\\

%         \vfill

%         Supervised by Your Advisor

%         \vfill
%         Department of Physics \& Astronomy\\
%         % Arts, Sciences and Engineering\\
%         School of Arts and Sciences

%         \vfill

%         % \Large
%         University of Rochester\\
%         2023\\
%         \vspace{1.0cm}
        
%     \end{center}
    
% \end{titlepage}

%% Fancy title page

\begin{titlepage}

    \center
    
    \textsc{\huge Senior Thesis}
    
    \vspace{0.4in}
    
    \noindent\makebox[\linewidth]{\rule{\linewidth}{1.2pt}}
    \vspace{0.0in} \\
    \textbf{\textsc{ {\LARGE Environmental Effects 
on Galaxy Evolution}} }\\
    \vspace{0.12in}
    \noindent\makebox[\linewidth]{\rule{\linewidth}{1.2pt}}
    
    \vspace{0.2in}
    
    \large
    \begin{minipage}{0.48\textwidth}
        \begin{flushleft}
            \textit{Author:} \\
            \textsc{Ziqi Mu} \\
        \end{flushleft}
    \end{minipage}
    \begin{minipage}{0.48\textwidth}
        \begin{flushright}
        \textit{Advisor:} \\
        \textsc{Prof. Segev BenZvi} \\
        \end{flushright}
    \end{minipage}
    
    \vspace{-0.4in}
    
    \begin{center}
       \includegraphics[width=0.9\textwidth]{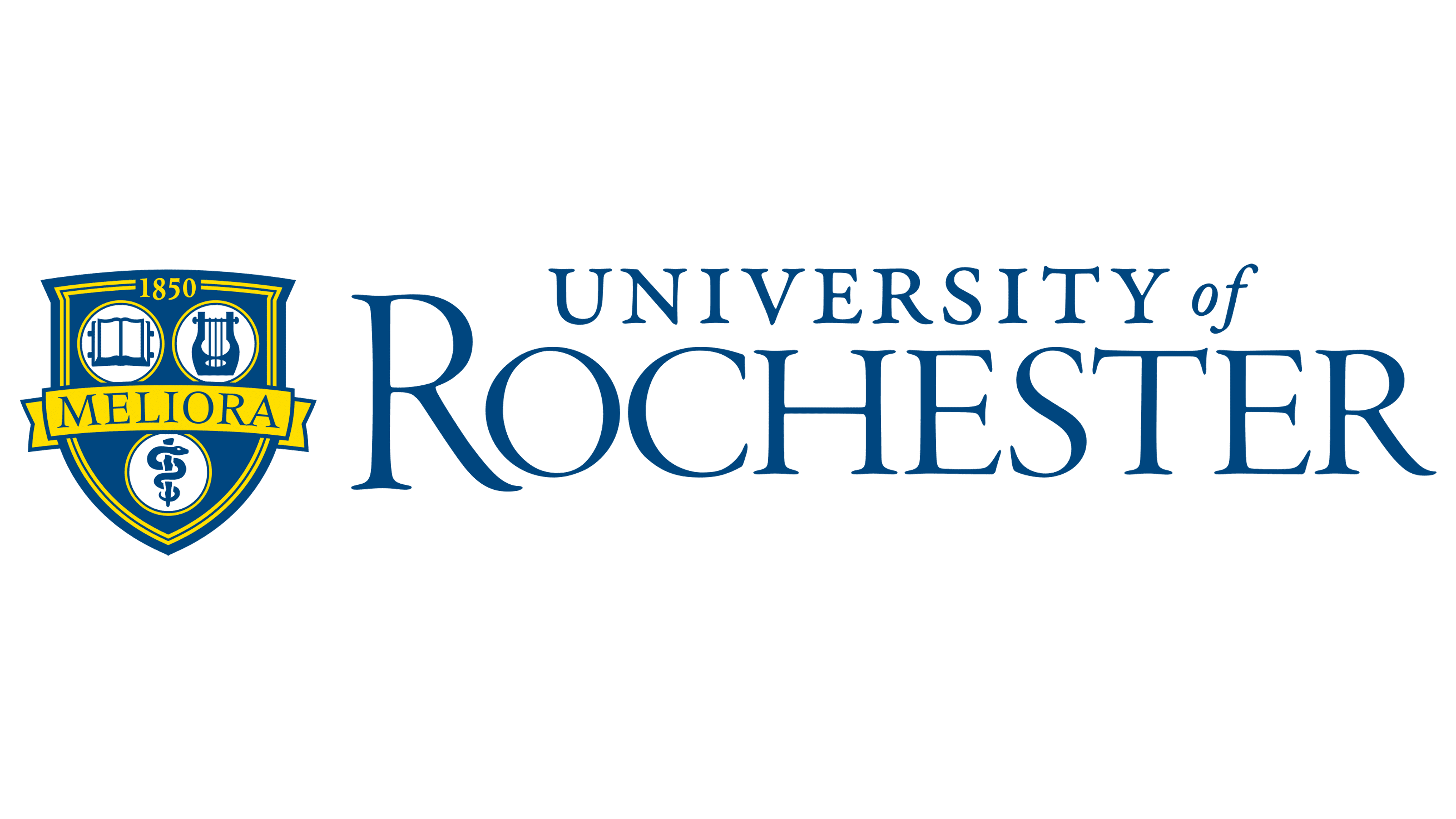} 
    \end{center}
    % \vspace{0.3in}
    
    \vspace{-0.6in}
    
    % \Large
    \textit{A senior thesis submitted to gain a fuller enrichment in the degree of}
    
    \vspace{0.3in}
    
    % \Large
    \textsc{Bachelor of Science \\
    Physics}
    
    \vspace{0.4in}
    
    % \large
    Department of Physics and Astronomy \\
    School of Arts and Sciences
    
    \vspace{0.4in}
    
    University of Rochester \\
    Spring \textsc{2025}

\end{titlepage}

\doublespacing

\chapter*{\centerline{Acknowledgements}}
\addcontentsline{toc}{chapter}{\numberline{}Acknowledgements}

I would like to express my sincere appreciation to my thesis advisor, Professor Segev BenZvi, for his exceptional guidance, support, and encouragement throughout this project. His mentorship has been invaluable, and I am truly thankful for the opportunity to learn from him.
I am also grateful to Professor Kelly Douglass for her helpful feedback and insightful suggestions, which contributed significantly to the progress of my work.  I’d also like to thank Hernan Rincon for his guidance and assistance, which were instrumental in overcoming key technical challenges.
I would like to express my deep appreciation to my family for their constant support and encouragement throughout my academic journey. I am also deeply thankful to my friends Wenbo Zhang, Yunwei Ni, Yiqian Jin, and Siqi Qin for their friendship, motivation, and the many moments of laughter and strength that helped me navigate this process.
Lastly, I am especially grateful to my cats, Toffee and Guppy, whose comforting presence and companionship helped me through a stressful period of life.

\chapter*{\centerline{Abstract}}
\addcontentsline{toc}{chapter}{\numberline{}Abstract}

\begin{center}
    \Large
    \vspace{-0.6cm}
    \textbf{\textsc{ {Environmental Effects 
on Galaxy Evolution }} }\\
        
    % \vspace{0.4cm}

    % \large
    % by

    % \Large
    % \vspace{0.2cm}
    % \textsc{Thomas Ahrens}
\end{center}

Galaxies evolve within a web-like cosmic structure, and their properties are strongly shaped by their surrounding environments. We apply a nonparametric Bayesian two-sample test based on Pólya tree priors to galaxy data from SDSS DR7 and DESI DR1 BGS to quantify the differences between galaxies in dense and sparse cosmic environments. Compared to the Kolmogorov-Smirnov test and parametric Bayesian test, our approach does not require strong assumptions about the underlying distributional form and provides a more sensitive and robust comparison of galaxy evolution across different environments. In particular, galaxies in VoidFinder voids tend to be fainter, less massive, and more star-forming compared to wall galaxies, while such contrasts are diminished under the $V^2$ REVOLVER pruning. These findings underscore the importance of both the statistical framework and the void classification algorithm in interpreting environmental effects on galaxy evolution.

\tableofcontents
\thispagestyle{plain}

\addcontentsline{toc}{chapter}{\numberline{}List of Figures}
\listoffigures

\addcontentsline{toc}{chapter}{\numberline{}List of Tables}
\listoftables

\clearpage
\pagenumbering{arabic}
\pagestyle{fancy}

\chapter{Introduction}

Galaxies in the universe are distributed in a web-like structure \cite{bond1996filaments} and they are constantly being
influenced by their surroundings. This cosmic structure sets the stage for one foundational question: How do galaxy environments affect their evolution?
For decades, scientists have been delving into this question. We are especially interested in studying galaxies in different regions of the universe: in dense regions (“walls”) where
galaxies interact, causing accelerated evolution; and in sparse regions (“voids”), where galaxies
evolve in relative isolation and are thought to have delayed evolution.

Studying galaxies in these
populations provides critical hints about the processes that drive their evolution.
Although we cannot measure galaxy evolution directly, we can observe galaxies’ properties and
use those as proxies for their evolutionary history. By dividing galaxies into different populations and searching for systematic differences in these properties
between different populations, we can analyze the evolution of galaxies.
To classify galaxies into populations of “voids” and “walls,” we use two popular void-finding algorithms: $V^2$ using REVOLVER pruning, a watershed algorithm that grows voids in underdense regions \cite{douglass2022vast}, and VoidFinder, an algorithm that grows and merges spherical void regions between galaxies \cite{douglass2022vast}. Both algorithms work on 3D galaxy positions from galaxy redshift surveys. We then classify galaxies as belonging to voids or walls and study the observed properties of the two populations.

\section{Galaxy Properties}

To investigate how the environment influences galaxy evolution, we mainly focus on six observable galaxy properties. By comparing these properties across galaxies located in voids and walls, we aim to understand how the environment affect stellar growth and star formation. The galaxy properties studied in our work include:

\begin{itemize}
    \item \textbf{Stellar mass} [\boldmath$\log(M_*/M_\odot)$]: the total mass contained in stars within a galaxy. Stellar mass increases over time as gas is converted into stars. In general, lower-density regions host proportionally a larger number of low-mass galaxies than higher-density ones \cite{vulcani2013galaxy}. Galaxies in wall environments tend to be more massive due to frequent interactions, whereas void galaxies typically evolve more passively and gradually accumulate stellar mass over time.

    \item \textbf{Absolute Magnitude ($M_r$)}: a measure of intrinsic brightness in the red (r-band) light. Since a galaxy’s absolute magnitude is closely linked to its stellar mass, brighter galaxies are generally more massive. Thus void galaxies are expected to be systematically fainter than galaxies in denser regions \cite{zaidouni2024impact}.

    \item \textbf{Color Indices ($g - r$, $u - r$)}: differences in magnitudes between photometric bands. Specifically, $u-r$ compares the ultraviolet (u-band) and red (r-band) light, while $g-r$ compares green (g-band) and red (r-band) light. Bluer colors are a result of hot O
and B stars \cite{zaidouni2024impact}, indicating younger, star-forming galaxies, while redder colors correspond to older, quiescent populations. Void galaxies are expected to be bluer than galaxies in denser regions.

    \item \textbf{Star Formation Rate} [\boldmath$\log(M_\odot/\mathrm{yr})$]: the rate of new star formation within a galaxy, typically measured in solar masses per year. Void galaxies are expected to be undergoing more star formation today than galaxies in denser regions,
due to potential retarded star formation and their continued presence of cold gas reservoirs \cite{zaidouni2024impact}. Void galaxies tend to form stars more gradually and continuously, therefore they tend to have higher star formation rates (SFR).

    \item \textbf{Specific Star Formation Rate} [\boldmath$\log(\mathrm{yr}^{-1})$]: the star formation rate per unit mass. This quantity captures the relative star-forming activity, independent of galaxy size. Void galaxies tend to exhibit higher specific star formation rates (sSFR).
\end{itemize}

\begin{figure}%[h]
  \centering
  \begin{subfigure}[t]{0.45\textwidth}
    \includegraphics[width=\linewidth]{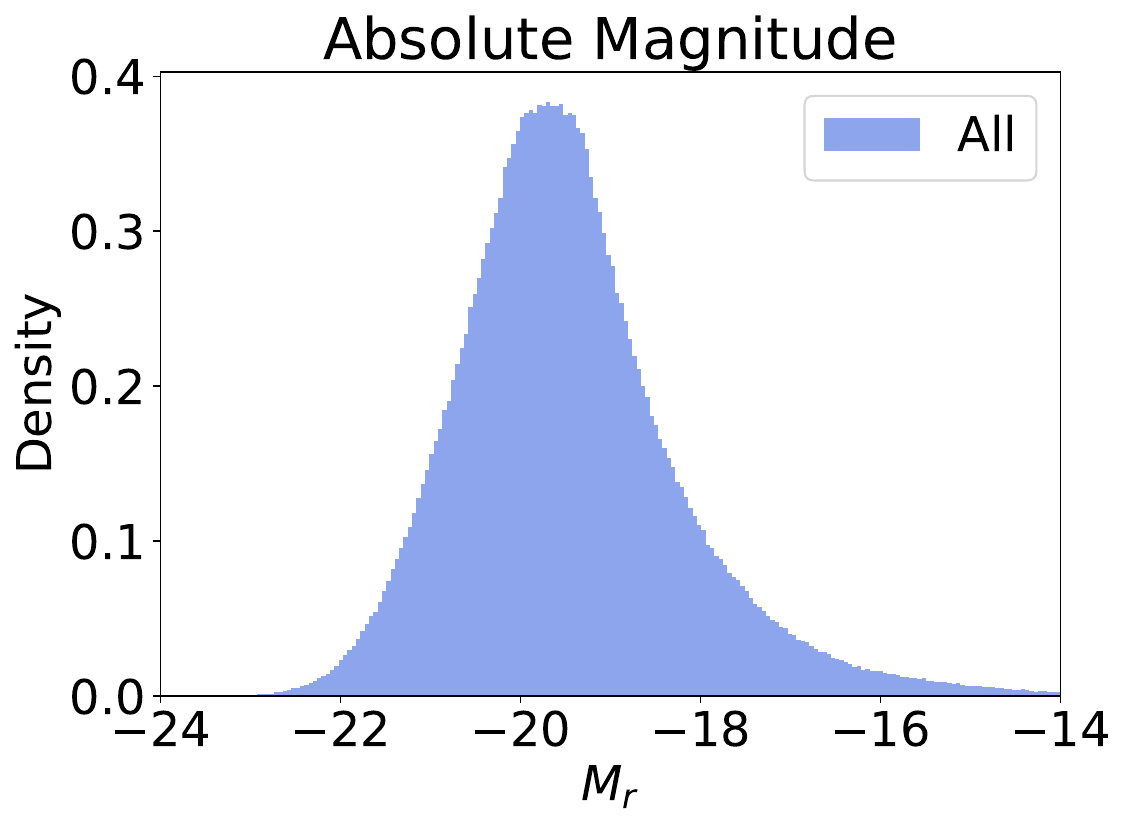}
  \end{subfigure}
  \hfill
  \begin{subfigure}[t]{0.45\textwidth}
    \includegraphics[width=\linewidth]{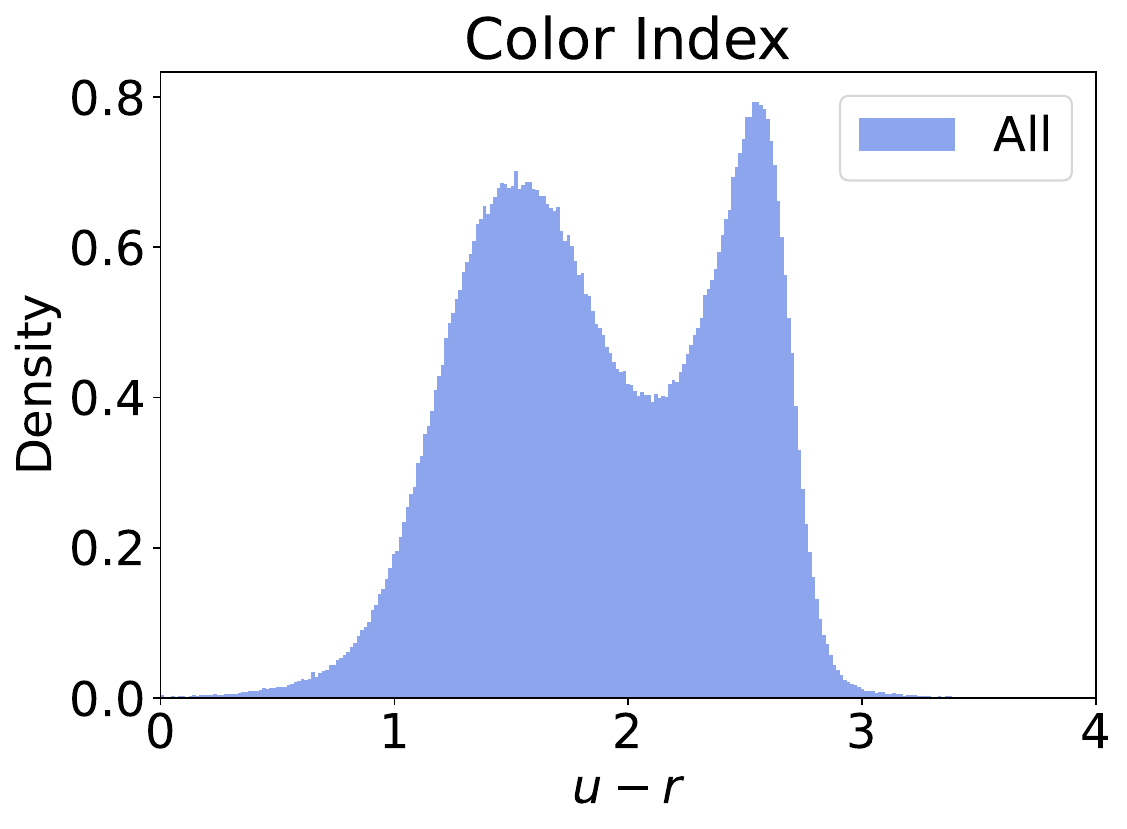}
  \end{subfigure}

  \caption[DESI BGS distributions of absolute magnitude and color]{Distribution of absolute magnitude (left) and color $u-r$ (right) for the full DESI BGS galaxy sample.}
  \label{fig:compare}
\end{figure}

Figure~\ref{fig:compare} illustrates the distribution of two representative galaxy properties: absolute magnitude ($M_r$) and color index ($u - r$) for the full galaxy sample. The histogram of $M_r$ shows a unimodal distribution, with most galaxies clustered around $M_r \approx -20$, reflecting the dominance of moderately luminous galaxies. In contrast, the $u - r$ color index exhibits a bimodal structure: the left peak corresponds to younger, bluer galaxies with ongoing star formation, while the right peak corresponds to older, redder galaxies that have ceased forming new stars.

These six properties offer a multi-dimensional view of galaxy evolution: void galaxies are expected to be fainter, less massive, bluer, and more star-forming than wall galaxies. The contrast between the two populations allows us to assess the role of the environment in shaping galaxy properties.

\section{Statistical Methods and Limitations}

To assess the statistical significance of differences in galaxy properties between void and wall regions, a very commonly used two-sample frequentist test is the Kolmogorov-Smirnov (KS) test. The
KS test uses the distance between the cumulative distribution functions (CDFs) of two observed populations to assess if they come from the same parent distribution. However, it has severe limitations in large datasets: the KS test only gives a binary decision (reject or fail to reject the null hypothesis, $H_0$). It does not tell us how much the distributions differ. Moreover, the KS test can yield an extremely low $p$-value since it computes the tail probability of the null hypothesis being true, thus leading to a false positive result (Type I error) \cite{filion2015signed}. Even if the two galaxy populations are only slightly different, the KS test will always reject $H_0$ due to the huge sample size.

Therefore, rather than relying on frequentist hypothesis testing, we adopt a Bayesian approach that directly quantifies the relative support for the null and alternative hypotheses. Specifically, we consider the ratio:
\begin{equation}
    \frac{\text{Pr}(y^{(1,2)} | H_0)}{\text{Pr}(y^{(1)}, y^{(2)} | H_1)} 
\end{equation}
where the numerator represents the marginal likelihood of observing the data under the null hypothesis $H_0$ (that the two populations are drawn from the same distribution), and the denominator represents the marginal likelihood under the alternative hypothesis $H_1$ (that the two populations are drawn from different distributions).

A detailed discussion of this Bayesian framework, including sensitivity
analysis and hyperparameter selection, will be presented in Chapter~2. 
The rest of the thesis is organized as follows: Section~3 describes the galaxy datasets and void identification algorithms used in our study. Section~4 presents the empirical results, comparing Bayesian methods and void classifications across various galaxy properties. Finally, Section~5 concludes and outlines potential directions for future research.

\chapter{Bayesian Nonparametric Testing Framework}
\label{chap:background}

Bayesian hypothesis testing compares competing hypotheses by directly quantifying the relative support that the observed data provide for each. The Bayes factor, defined as the ratio of the marginal likelihoods under the null and alternative hypotheses, serves as a key measure in this framework. Unlike traditional frequentist testing, which yields a binary reject-or-not decision \cite{hoijtink2019tutorial} based on $p$-values, the Bayes factor enables direct comparison and interpretation of the strength of evidence in favor of one hypothesis over the other.

In previous work by Zaidouni et al.\ \cite{zaidouni2024impact}, a parametric Bayesian framework was used to test for differences in galaxy properties between void and wall environments. Their model assumes that the observed data can be well-described by one or more Gaussian or skew-normal distributions (i.e., a mixture model). Under this assumption, the data can be binned and fit under two competing hypotheses: either both samples come from the same parent distribution ($H_0$), or from two different parent distributions ($H_1$). A Bayes factor is then used to compare these two models.

Although parametric Bayesian tests are easy to set up, they also suffer from major weaknesses. Since they require specifying the shape of the distribution beforehand, they can be overly restrictive if the true data-generating process deviates significantly from the assumed form. Furthermore, they are sensitive to outliers and can be significantly affected by extreme values.
To reduce our dependence on assuming a specific shape, we aim to apply a non-parametric two-sample Bayesian test. Compared to the parametric Bayes factor approach, the non-parametric Bayesian test has a more complex setup and is computationally demanding, but it provides a more robust and flexible analysis of the differences between void and wall galaxy distributions.  As first proposed in \cite{holmes2015two}, P\'olya tree priors can be used to derive the Bayes factor of the non-parametric test. It can accommodate known uncertainty in the form of the underlying
sampling distribution and provides an explicit posterior probability measure of both dependence
and independence \cite{filippi2017bayesian}.

\section{P\'olya Tree Priors}

\begin{figure}%[H]
    \centering
    \includegraphics[scale=0.22]{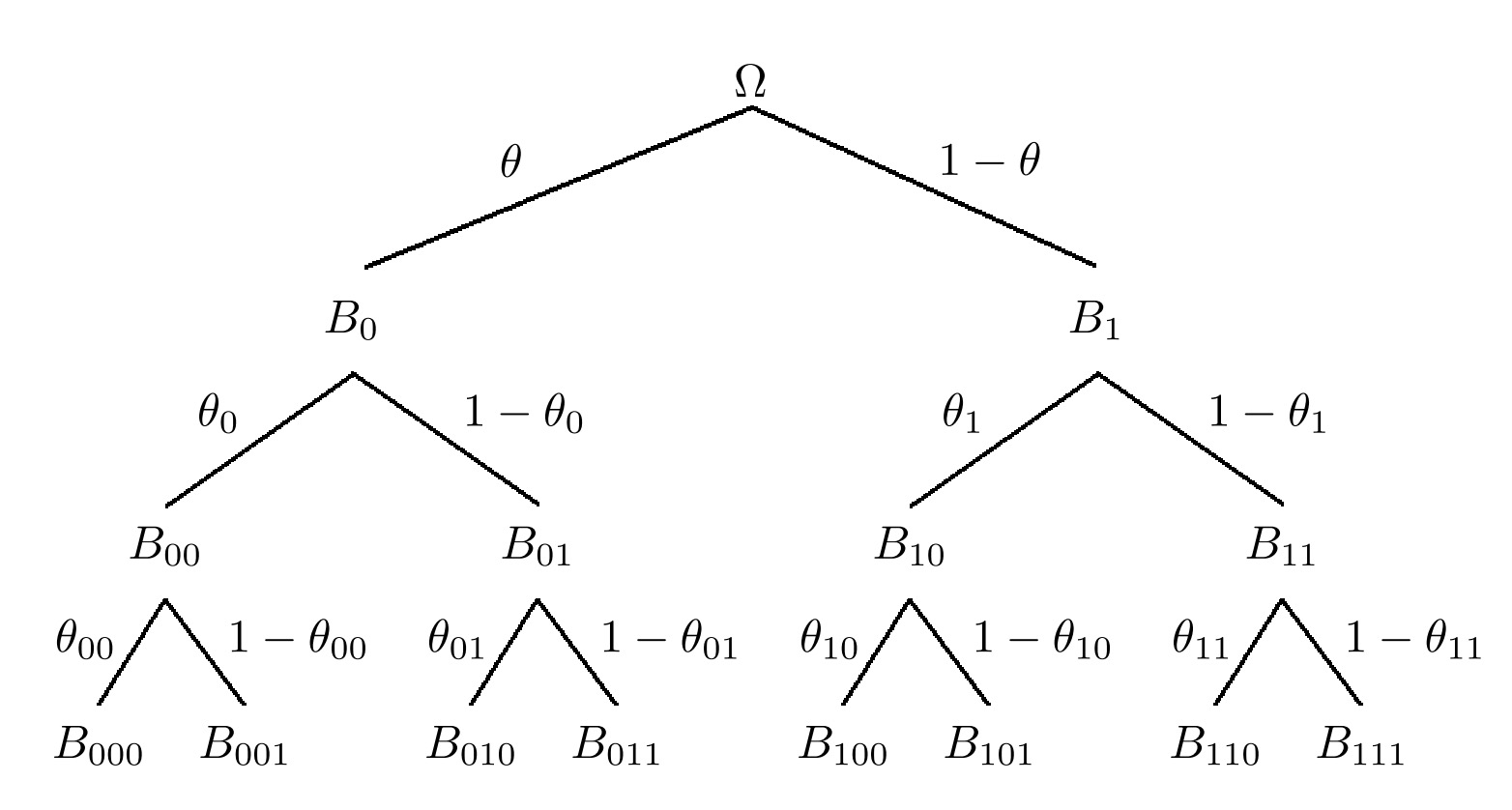}
    \caption[Pólya tree distribution]{Construction of a Pólya tree distribution, each of the $\theta_{\epsilon_m}$ is independently drawn from $\text{Beta}(\alpha_{\epsilon_m 0}, \alpha_{\epsilon_m 1})$ \cite{holmes2015two}. Adapted from Ferguson \cite{ferguson1974prior}. }
    \label{fig:polya_tree}
\end{figure}

To flexibly model the distribution of galaxy properties without assuming a fixed parametric form, we adopt a \text{Pólya tree prior} --- a Bayesian nonparametric prior over probability distributions \( F \) on a domain \(\Omega = [0, 1)\). A Pólya tree recursively partitions \( \Omega \) into binary subsets, forming a bifurcating tree structure \cite{holmes2015two}. As illustrated in Figure~\ref{fig:polya_tree}, at each node \( j \), a branching probability \( \theta_j \sim \text{Beta}(\alpha_{j0}, \alpha_{j1}) \) \cite{lavine1992some} determines how probability mass is split between the left and right branches. The probability of falling into a particular subset \( B_{\boldsymbol{\epsilon}_k} \) is given by the product of branching probabilities along the unique path
\begin{equation}
P(B_{\boldsymbol{\epsilon}_k}) = \prod_{i=1}^{k} 
\left( \theta_{\boldsymbol{\epsilon}_{i-1}} \right)^{1 - \epsilon_{ii}}
\left( 1 - \theta_{\boldsymbol{\epsilon}_{i-1}} \right)^{\epsilon_{ii}}
\end{equation}
where \( \boldsymbol{\epsilon}_k = \{ \epsilon_1, \epsilon_2, \ldots, \epsilon_k \} \) is a binary vector representing the sample path of the particle down to level \( k \), with each element \( \epsilon_i \in \{0, 1\} \) indicating a left (\( 0 \)) or right (\( 1 \)) branch taken at level \( i \).
The \( \theta_j \) are mutually independent, and the prior is conjugate: after observing data, the posterior remains a Pólya tree with updated parameters.

In practical applications, the tree is truncated at some finite depth \( m \), resulting in a partially specified Pólya tree. According to Holmes \cite{holmes2015two}, at each level \( m \), we set \( \alpha_{m0} = \alpha_{m1} = c_m \) for some level-dependent constant \( c_m \), where the setting of \( c_m \) governs the underlying continuity of the resulting $F$’s.
Following the framework introduced by Ferguson et al. \cite{ferguson1974prior}, we set $c_m = c m^2$ with $c > 0$, ensuring that the sampled measure $F$ is absolutely continuous with probability 1.
The value of the precision parameter \( c \) can be chosen to control the rate at which the variance decreases across tree levels and will be discussed in the later section. 

Let $\Pi$ denote the collection of sets $\{B_0, B_1, B_{00}, \ldots\}$ and let ${A}$ denote the collection of parameters that determine the distribution at each junction, ${A} = (\alpha_{00}, \alpha_{01}, \alpha_{000}, \ldots)$ \cite{holmes2015two}. Given data \( y \) and Beta-distributed branching variables \( \theta_j \), we can marginalize out \( \theta_j \) using the Beta-Binomial identity, yielding:
\begin{equation}
\Pr(y \mid \Pi, A) = \prod_j 
\left[
\frac{\Gamma(\alpha_{j0} + \alpha_{j1})}{\Gamma(\alpha_{j0}) \Gamma(\alpha_{j1})}
\frac{\Gamma(\alpha_{j0} + n_{j0}) \Gamma(\alpha_{j1} + n_{j1})}
{\Gamma(\alpha_{j0} + \alpha_{j1} + n_{j0} + n_{j1})}
\right]
\end{equation}
where \( n_{j0} \) and \( n_{j1} \) denote the number of observations falling into the left and right children of node \( j \), respectively. The product runs over all partitions in the tree up to depth \( m \). This structure enables efficient evaluation of model evidence, which we will later use for hypothesis testing.

\section{Bayesian Hypothesis Testing with Pólya Trees}

Unlike the classical (frequentist) approach, which treats the population parameter as a fixed but unknown value and relies on repeated sampling to derive estimates, the Bayesian approach treats the parameter as a random variable with a probability distribution. In the frequentist framework, uncertainty comes from the randomness of the sample, and repeated sampling is used to construct confidence intervals or perform hypothesis testing. In contrast, Bayesian inference assigns a prior distribution to the parameter and updates it using observed data to obtain a posterior distribution via Bayes' theorem.

Given a parameter $\theta$ and data $y$, Bayesian inference updates prior knowledge $\text{Pr}(\theta)$ using the likelihood $\text{Pr}(y|\theta)$ to obtain the posterior distribution:
\begin{equation}
\text{Pr}(\theta | y) = \frac{\text{Pr}(y | \theta) \text{Pr}(\theta)}{\text{Pr}(y)}
\end{equation}
where \text{Pr}(y) integrates out the dependence on $\theta$:
\begin{equation}
    \text{Pr}(y) = \int \text{Pr}(y \mid \theta) \text{Pr}(\theta) \, d\theta.
\end{equation}
Since $\text{Pr}(y)$ contains no information regarding $\theta$, we can say that it is a normalizing constant \cite{hoff2009first}, and the unnormalized posterior is proportional to the likelihood times prior:
\begin{equation}
    \text{Pr}(\theta | y) \propto \text{Pr}(y | \theta) \text{Pr}(\theta).
\end{equation}

Since we want to quantify the difference between void galaxies and wall galaxies, we are interested in providing a weight of evidence in favor of $H_{0}$ given the observed data \cite{holmes2015two}. From Bayes theorem,
\begin{equation}
\text{Pr}(H_0 | y^{(1,2)}) \propto \text{Pr}(y^{(1,2)} | H_0) \text{Pr}(H_0).
\end{equation} 
where \( \text{Pr}(H_0) \) is the prior probability of the null hypothesis \( H_0 \), \( \text{Pr}(y^{(1,2)} | H_0) \) is the likelihood of observing $y^{(1,2)}$ under \( H_0 \), and  $y^{(1)}$ and $y^{(2)}$ are samples from some common distribution $F^{(1,2)}$ with $F^{(1,2)}$ unknown. According to Holmes \cite{holmes2015two}, a Pólya tree prior is adopted to specify uncertainty in \( F^{(1,2)} \):
\begin{equation}
    F^{(1,2)} \sim \text{PT}(\Pi, A).
\end{equation}
Under the alternative hypothesis \( H_1 \), we assume independent distributions for \( y^{(1)} \sim F^{(1)} \) and \( y^{(2)} \sim F^{(2)} \), where \( F^{(1)} \) and \( F^{(2)} \) are unknown. Again, we adopt a Pólya tree prior for these distributions with the same prior parameterization:
\begin{equation}
    F^{(1)}, F^{(2)}, F^{(1,2)} \overset{\text{iid}}{\sim} \text{PT}(\Pi, A).
\end{equation}

The posterior odds on \( H_0 \) can be written as:
\begin{equation}
    \frac{\text{Pr}(H_0 | y^{(1,2)})}{\text{Pr}(H_1 | y^{(1)}, y^{(2)})} =
    \frac{\text{Pr}(y^{(1,2)} | H_0)}{\text{Pr}(y^{(1)}, y^{(2)} | H_1)}
    \frac{\text{Pr}(H_0)}{\text{Pr}(H_1)}.
\end{equation}

The Bayes factor $B_{01}$ is defined as the ratio of marginal likelihoods and quantifies how much the data updates our belief in one hypothesis relative to the other \cite{Kass01061995}:
\begin{equation}
    B_{01} = \frac{\text{Pr}(y^{(1,2)} | H_0)}{\text{Pr}(y^{(1)}, y^{(2)} | H_1)} = \prod_{j} b_j.
\end{equation}
where
\begin{equation}
\begin{aligned}
    b_j &=
    \frac{\Gamma(\alpha_{j0}) \Gamma(\alpha_{j1})}{\Gamma(\alpha_{j0} + \alpha_{j1})}
    \cdot
    \frac{\Gamma(\alpha_{j0} + n^{(1)}_{j0} + n^{(2)}_{j0}) \Gamma(\alpha_{j1} + n^{(1)}_{j1} + n^{(2)}_{j1})}
    {\Gamma(\alpha_{j0} + n^{(1)}_{j0} + n^{(2)}_{j0} + \alpha_{j1} + n^{(1)}_{j1} + n^{(2)}_{j1})} \\[10pt]
    &\quad \times
    \frac{\Gamma(\alpha_{j0} + n^{(1)}_{j0} + \alpha_{j1} + n^{(1)}_{j1})}
    {\Gamma(\alpha_{j0} + n^{(1)}_{j0}) \Gamma(\alpha_{j1} + n^{(1)}_{j1})}
    \cdot
    \frac{\Gamma(\alpha_{j0} + n^{(2)}_{j0} + \alpha_{j1} + n^{(2)}_{j1})}
    {\Gamma(\alpha_{j0} + n^{(2)}_{j0}) \Gamma(\alpha_{j1} + n^{(2)}_{j1})}.
\end{aligned}
\end{equation}
The quantities \( n^{(1)}_{j0} \) and \( n^{(1)}_{j1} \) represent the numbers of observations in \( y^{(1)} \) that fall to the right and left at each junction, respectively. Similarly, \( n^{(2)}_{j0} \) and \( n^{(2)}_{j1} \) are the corresponding quantities for \( y^{(2)} \) \cite{holmes2015two}.

We use the Bayes factor criteria adapted from \cite{Kass01061995}:

\begin{table}[H]
\centering
\caption[Interpretation of Bayes factors]{Interpretation of Bayes Factors}
\label{tab:bayes_factor_interpretation}
\begin{tabular}{ccc}
\toprule
\textbf{Value of \( B_{01} \)} & \textbf{Value of \( \log(B_{01}) \)} & \textbf{Evidence Against \( H_0 \)} \\
\midrule
1 to 0.32       & 0 to -1/2              & Not worth more than a bare mention \\
0.32 to 0.1     & -1/2 to -1             & Substantial                         \\
0.1 to 0.01     & -1 to -2           & Strong                              \\
\(< 0.01\)      & \(<-2\)           & Decisive                            \\
\bottomrule
\end{tabular}
\end{table}

\section{Sensitivity Analysis and Hyperparameter Choice}

To validate our implementation of the Bayesian two-sample test based on the Pólya tree prior, we replicate three of the canonical simulation settings described in Holmes \cite{holmes2015two}. These settings are designed to evaluate the sensitivity of the Bayes factor under both the null hypothesis and controlled departures from it.
\begin{enumerate}[label=\text{\alph*)}]
    \item \text{Null:} Both samples are drawn from the same distribution: \( Y^{(1)} \sim \mathcal{N}(0, 1) \), \( Y^{(2)} \sim \mathcal{N}(0, 1) \), representing the case with no difference.
    
    \item \text{Mean shift:} Sample 1 is generated from \( Y^{(1)} \sim \mathcal{N}(0, 1) \), while Sample 2 is generated from \( Y^{(2)} \sim \mathcal{N}(1, 1) \), introducing a difference in mean but not variance.
    
    \item \text{Variance shift:} Sample 1 is drawn from \( Y^{(1)} \sim \mathcal{N}(0, 1) \), and Sample 2 from \( Y^{(2)} \sim \mathcal{N}(0, 4) \), introducing a difference in variance but not in mean.
\end{enumerate}

We set the Pólya Tree parameters $\alpha_m = c_m = c m^2$ with $c > 0$, ensuring that the sampled measure $F$ is absolutely continuous with probability 1, consistent with the setup described in Holmes ~\cite{holmes2015two} and following the framework introduced by Ferguson et al.~\cite{ferguson1974prior}.

We follow the sample size settings used in Holmes' study, with \( n = 10, 50, 100, 200 \) and 500 replications per configuration based on the PTTests implementation by Boeken \cite{boekenPTTests}. While Holmes reports the mean Bayes factor across replications, we instead summarize the results using the median and the 16th–84th percentile interval to better reflect the skewness and variability in the Bayes factor distribution.

\begin{figure}%[t]
    \centering
    \includegraphics[scale=0.45]{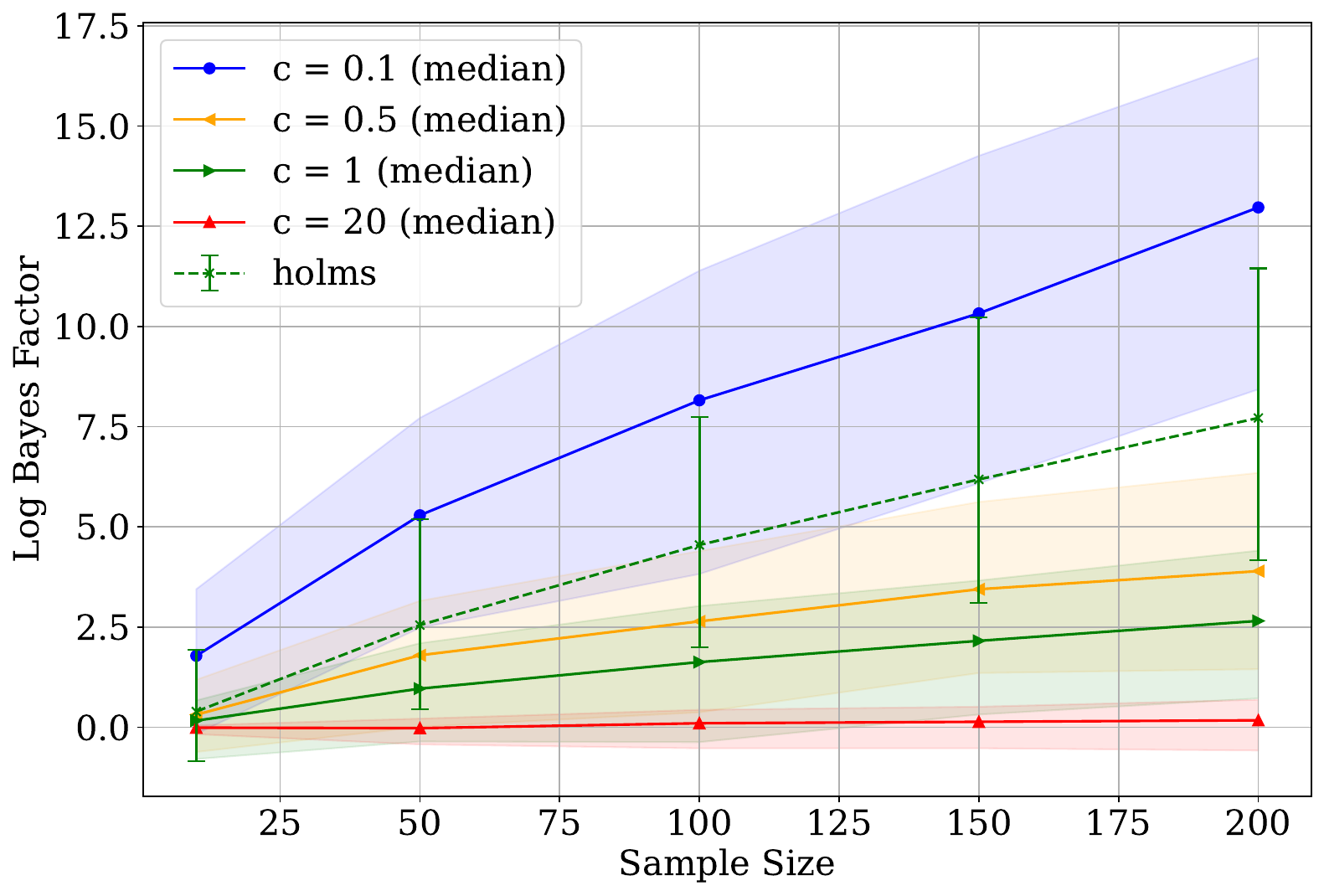}
    \caption[Median log Bayes factor versus sample size for the null scenario]{Median log Bayes factor with respect to the sample size under the null scenario. The shaded regions represent the 16th–84th percentile intervals over repeated simulations. The dashed green curve corresponds to the reference results from Holmes et al.~\cite{holmes2015two}}
    \label{fig:nullcase}
\end{figure}

\begin{figure}%[t]
    \centering
    \begin{subfigure}[b]{0.48\textwidth}
        \centering
        \includegraphics[width=\linewidth]{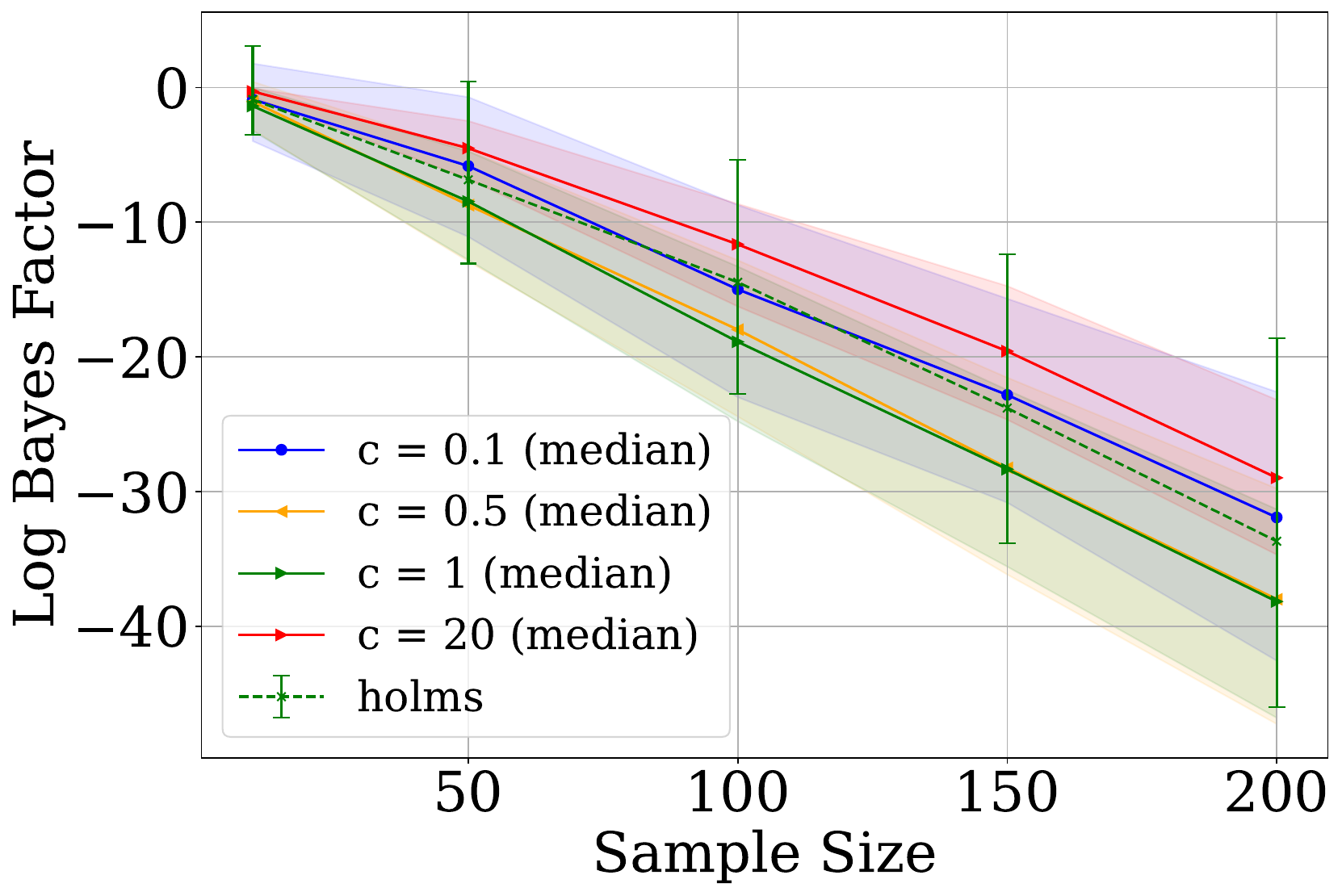}
        \caption{Alternative: Mean shift}
        \label{fig:meanshift}
    \end{subfigure}
    \hspace{0.01\textwidth} 
    \begin{subfigure}[b]{0.48\textwidth}
        \centering
        \includegraphics[width=\linewidth]{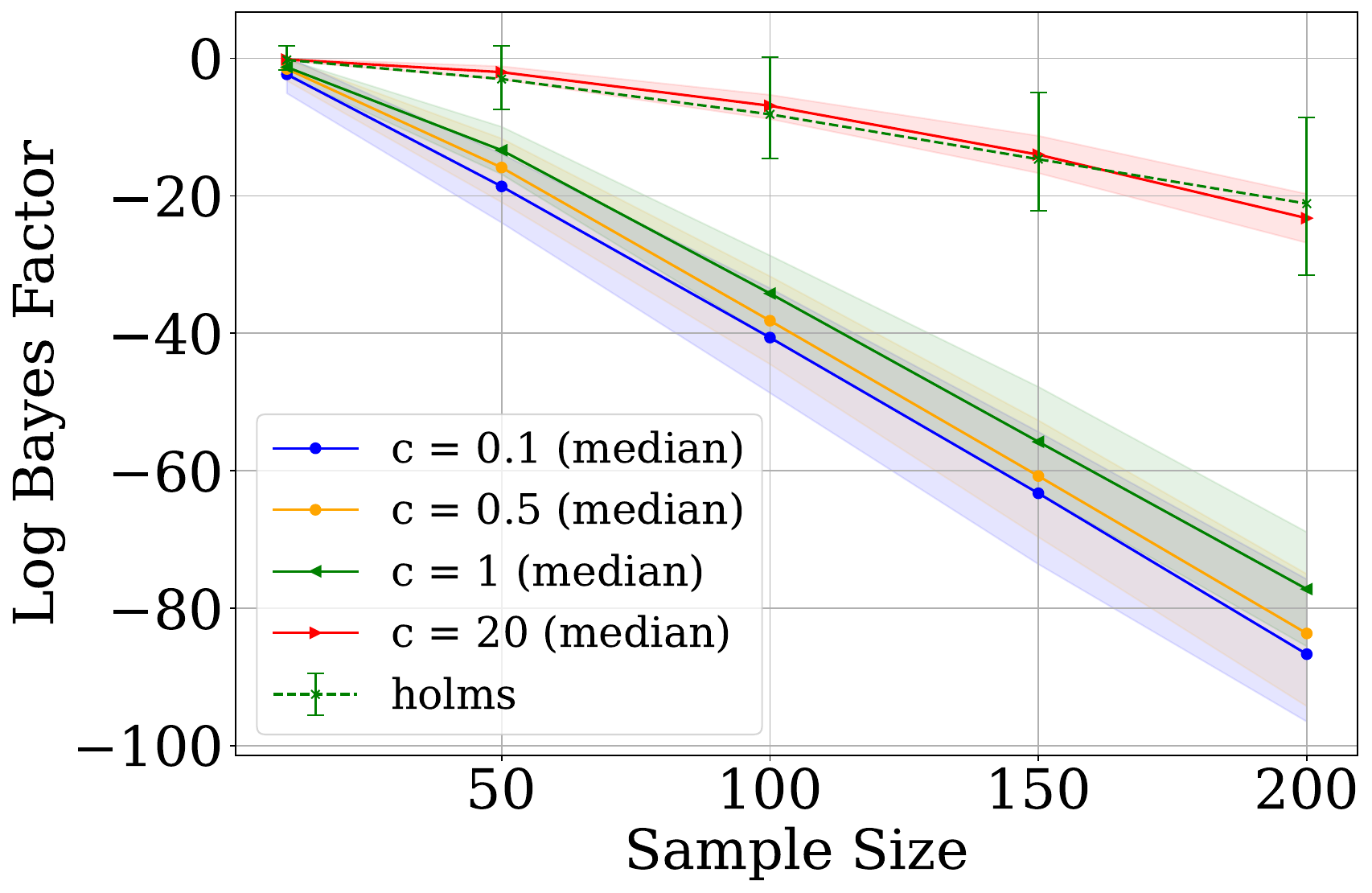}
        \caption{Alternative: Variance shift}
        \label{fig:varshift}
    \end{subfigure}
    \caption[Bayes factor versus sample size for different means and variances]{Median log Bayes factor with respect to the sample
size under two alternative scenarios. (a) Samples differ in mean. (b) Samples differ in variance. The shaded regions represent the 16th–84th percentile intervals over repeated simulations. The dashed green curve corresponds to the reference results from Holmes et al.~\cite{holmes2015two}.
}
    \label{fig:alt_shift}
\end{figure}

The results reveal that the precision parameter \( c \) plays a pivotal role in shaping the sensitivity of the Bayes factor across different testing scenarios. As illustrated in Figures~\ref{fig:nullcase} and \ref{fig:alt_shift}, both the value of \( c \) and the sample size substantially influence the test’s behavior under the null and alternative hypotheses.

According to Figure~\ref{fig:nullcase}, under the null hypothesis, the log Bayes factor remains consistently positive across all values of \( c \), indicating support for \( H_0 \). As the sample size increases, the log Bayes factor becomes more strongly positive, reflecting increased certainty in favor of the null.  Importantly, larger values of \( c \) yield more conservative and stable estimates of the Bayes factor, while smaller values produce higher variability across replications. This behavior is theoretically grounded in the structure of the Pólya tree prior: the prior variance of each partition probability \( P(B_{\epsilon_k}) \) diminishes and tends to zero as \( c \rightarrow \infty \), leading to stronger shrinkage of the posterior toward the centering distribution \cite{hanson2006inference}. 
Consequently, large values of \( c \) help stabilize the posterior but may reduce responsiveness to minor sample fluctuations.

In the mean shift scenario (Figure~\ref{fig:meanshift}), small values of \( c \) result in increasingly negative log Bayes factors as sample size grows, correctly identifying the distributional shift. However, for large values of \( c \) (e.g., \( c = 20 \)), the test becomes less sensitive: the log Bayes factor curve flattens and shifts up toward zero, even with a substantial sample size. This is consistent with the theoretical result that the expected squared Euclidean distance between two independently drawn Pólya tree distributions decreases with increasing \( c \) \cite{holmes2015two}. 
Thus, excessively large values of \( c \) lead to overly similar priors under the alternative, diminishing the test’s ability to detect real mean differences, even as the sample size increases.

In the variance shift setting (Figure~\ref{fig:varshift}), this loss of sensitivity becomes even more pronounced. While our experiments fix the tree depth at \( m = 8 \), increasing \( c \) imposes stronger prior concentration across all levels, effectively suppressing the posterior’s ability to adapt to variation in distributional spread. As a result, the log Bayes factor curve shifts upward and flattens across all sample sizes, showing a decline in detection power as \( c \) increases.

These findings highlight the importance of considering both the precision parameter \( c \) and the sample size \( n \) when applying Pólya tree-based Bayesian two-sample tests. As discussed in Holmes \cite{holmes2015two}, values of \( c \) between 1 and 10 are generally effective in practice. Similarly, Hanson \cite{hanson2006inference} notes that when \(c=1 \), the resulting random densities often exhibit a high degree of flexibility, making the prior well-suited for modeling complex distributions. Thus we set \( c = 1 \) in all subsequent applications.

\chapter{Data and Void-Finding Algorithms}
\label{chap: Fitting Procedure}

\section{Data and Selection}

\subsection{SDSS DR7}

The Sloan Digital Sky Survey Data Release 7 (SDSS DR7) marks the completion of the original SDSS goals and the conclusion of the SDSS-II phase. This catalog includes a total of 11,663 deg$^2$ of imaging data, an increase of about 2,000 deg$^2$ over previous releases, and contains five-band photometry for approximately 357 million objects \cite{abazajian2009seventh}. In total, the SDSS DR7 sample includes more than 1.6 million spectra, including 930,000 galaxies, 120,000 quasars, and 460,000 stars. It represents a significant milestone in astronomical surveys, providing a vast and detailed dataset for the scientific community. According to \cite{zaidouni2024impact}, we use SDSS DR7 as reprocessed and presented in the NASA-Sloan Atlas (NSA, version 1.0.1), a catalog of images and parameters derived from the SDSS imaging. 
The NSA provides \textit{K}-corrected absolute magnitudes, colors, and stellar masses. Star formation rates (SFR) and specific star formation rates (sSFR) are obtained from the MPA-JHU value-added galaxy catalog \cite{zaidouni2024impact}. According to the SDSS DR7 data flags, we keep galaxies within the main survey footprint \cite{void_analysis_bayesfactor}. To focus on the contiguous northern galactic cap, we further restrict right ascension to the range \(110^\circ < \mathrm{RA} < 270^\circ\). Additionally, we exclude a narrow, high-declination stripe (\(250^\circ < \mathrm{RA} < 269^\circ\) and \(51^\circ < \mathrm{DEC} < 67^\circ\)) that lies outside the main contiguous region, resulting in a clean and uniform subsample.

\subsection{DESI DR1 Bright Galaxy Survey}

The Dark Energy Spectroscopic Instrument (DESI) is an ongoing, state-of-the-art galaxy-redshift survey that will cover 14,000 deg$^2$ of the sky and measure the redshifts of over 40 million galaxies and quasars out to z $>$ 3 \cite{rincon2024desivast}. 
DESI targets include overlapping populations of nearby bright galaxies, luminous red galaxies, emission line galaxies, and quasars. To facilitate a complex survey with multiple target classes, DESI has extensive data collection and processing pipelines, including procedures for survey-tiling, spectroscopic reduction, and redshift-fitting.
A catalog from the DESI DR1 Bright Galaxy Survey (BGS) was selected and we follow the galaxy selection criteria outlined by Rincon et al. \cite{rincon2024desivast}. Galaxies brighter than \( M_r \leq -20 \) were selected, as this population is known to trace large-scale structure. With both $K$-corrections and $E$-corrections applied, the magnitude cut of $M_{r} \leq -20$ and the BGS Bright survey apparent magnitude limit of $m_r \leq 19.5$ correspond to a redshift limit of $z \leq 0.24$, which represents the maximum distance at which voids can be found in the BGS Bright survey.

Unlike SDSS DR7, the DESI DR1 BGS sample does not provide direct measurements of star formation rate or specific star formation rate. To address this limitation, we use the H$\alpha$ equivalent width (EW), which is defined as the ratio of H$\alpha$ flux (tracing instantaneous SF activity) and continuum flux density (tracer of stellar mass), as a proxy for specific star formation rate (sSFR) \cite{khostovan2021correlations}.

\section{Void-Finding Algorithms}
\label{sec:void-methods}

We identify cosmic voids using two widely adopted void-finding algorithms implemented in the Void Analysis Software Toolkit \cite{douglass2023updated}: the sphere-growing algorithm VoidFinder and the watershed-based ZOBOV algorithm \cite{rincon2024desivast}, here referred to as $V^2$.

\subsection{VoidFinder}

\begin{figure}%[htbp]
    \centering
    \includegraphics[width=\textwidth]{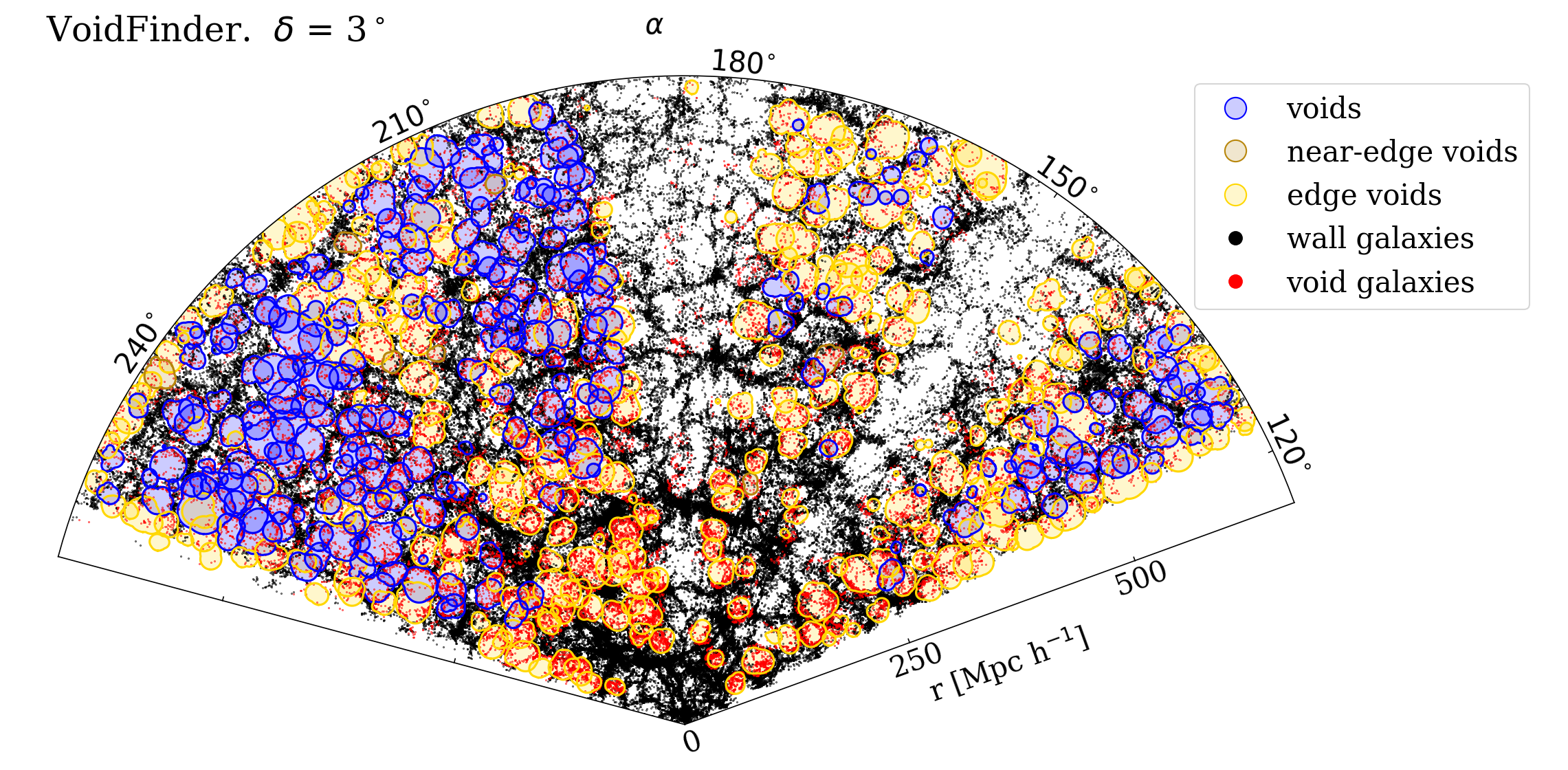}
    \caption[VoidFinder voids in DESI BGS]{VoidFinder slice plot for BGS galaxies at a declination of \( \delta = 3^\circ \). Void (wall) galaxies are shown in red (black). Edge (interior) voids are shown in yellow (blue). \cite{rincon2024desivast}}
    \label{fig:voidfinder}
\end{figure}

The VoidFinder algorithm operates on a volume-limited galaxy sample, classifying galaxies into highly clustered “wall” populations and isolated “field” populations based on their third-nearest neighbor distance \cite{rincon2024desivast}. Galaxies with a third-nearest neighbor beyond $5.33\,h^{-1}\,\mathrm{Mpc}$ \cite{zaidouni2024impact} are identified as field galaxies and temporarily removed to reveal underdense regions. The remaining wall galaxies are mapped onto a 3D grid (typically 5$h^{-1}$Mpc resolution). From the center of each empty grid cell, spheres are seeded and grown until their surfaces touch at least four wall galaxies. Spheres that are non-overlapping (less than $10\%$ volume overlap) and exceed 10$h^{-1}$Mpc in radius are marked as maximal spheres and define individual voids. All other spheres of a given void must overlap only their void’s maximal sphere by at least 50$\%$ of their volumes. Figure~\ref{fig:voidfinder} presents an example of void regions identified by the VoidFinder algorithm using data from DESI DR1 BGS \cite{vastgithub}. Galaxies that lie within the boundaries of these voids are classified as void galaxies, while those outside are considered wall galaxies.

\subsection{$V^2$ REVOLVER}

The $V^2$ algorithm defines voids based on the galaxy density field, which is estimated using a Voronoi tessellation of the galaxy distribution \cite{rincon2024desivast}. Each galaxy is associated with a Voronoi cell whose volume serves as a proxy for the local density. In $V^2$, adjacent low-density cells are grouped into structures called ``zones." The REVOLVER pruning method, a specific variant of the $V^2$ framework, defines voids by treating each zone independently without combining them. To ensure the reliability of void boundaries, any Voronoi cell that extends beyond the survey mask is assigned an infinite density, effectively confining all identified voids within the observed volume. 

\begin{figure}%[htbp]
    \centering
    \includegraphics[width=1.0\textwidth]{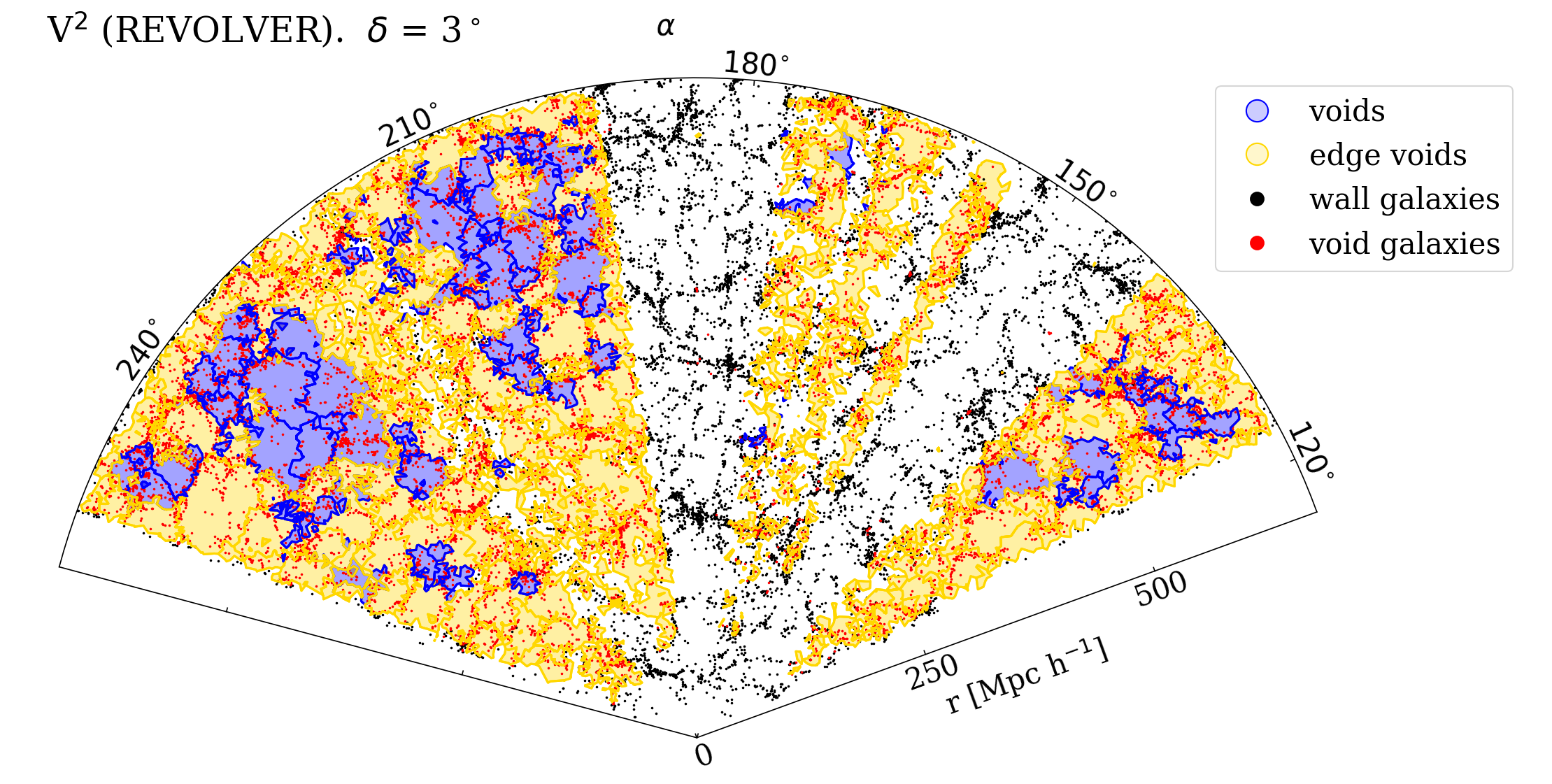}
    \caption[$V^2$ REVOLVER voids in DESI BGS]{$V^2$ REVOLVER slice plot for BGS galaxies at a declination of \( \delta = 3^\circ \). Void (wall) galaxies are shown in red (black). Edge (interior) voids are shown in yellow (blue). \cite{rincon2024desivast}}
    \label{fig:v2}
\end{figure}

While both VoidFinder and $V^2$ REVOLVER aim to identify void regions in the cosmic web, they differ in methodology and implementations. VoidFinder identifies dynamically-distinct regions with low shell-crossing numbers \cite{veyrat2023comparison} and tends to produce smaller, more numerous voids. In contrast, $V^2$ REVOLVER voids often include wall-like structures within the void volume and combine adjacent dynamically distinct void regions into single voids. Consequently, as Zaidouni et al. found \cite{zaidouni2024impact}, galaxies within $V^2$ REVOLVER voids do not exhibit strong differences from wall galaxies.
Figure~\ref{fig:v2} presents an example of void regions identified by the $V^2$ REVOLVER algorithm using data from DESI DR1 BGS \cite{vastgithub}.

\chapter{Results}
\label{chap:chapter-4}

\section{Comparison with Parametric Bayesian Method through SDSS DR7}

A key goal of our study is to assess the effectiveness of the nonparametric Bayesian model in detecting differences between void and wall galaxy populations. We first apply the nonparametric Bayesian test to SDSS DR7 data and compare its performance with the parametric Bayesian results adapted from \cite{zaidouni2024impact}, across various galaxy properties under the VoidFinder and $V^2$ REVOLVER void classifications. 

\begin{figure}%[h]
  \centering
  % First row
  \begin{subfigure}[t]{0.45\textwidth}
    \includegraphics[width=\linewidth]{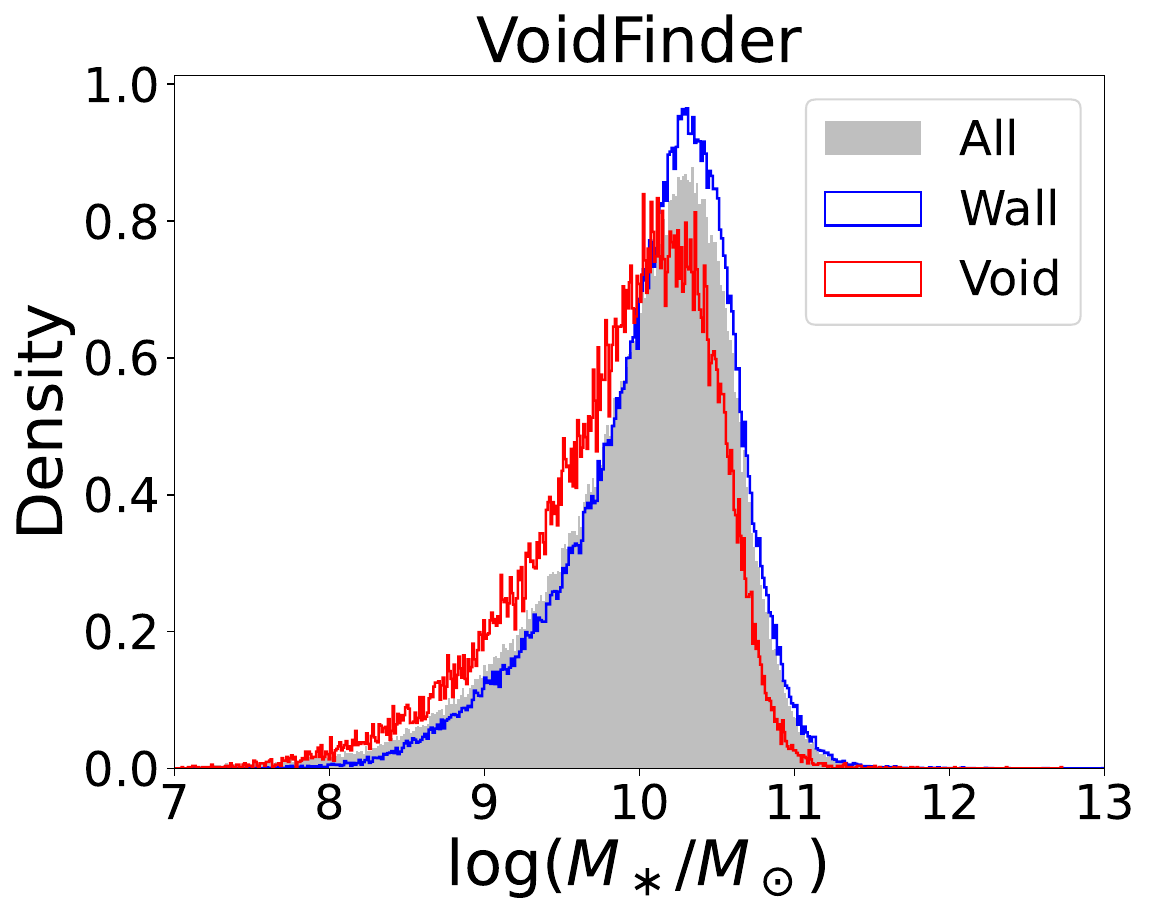}
  \end{subfigure}
  \hfill
  \begin{subfigure}[t]{0.45\textwidth}
    \includegraphics[width=\linewidth]{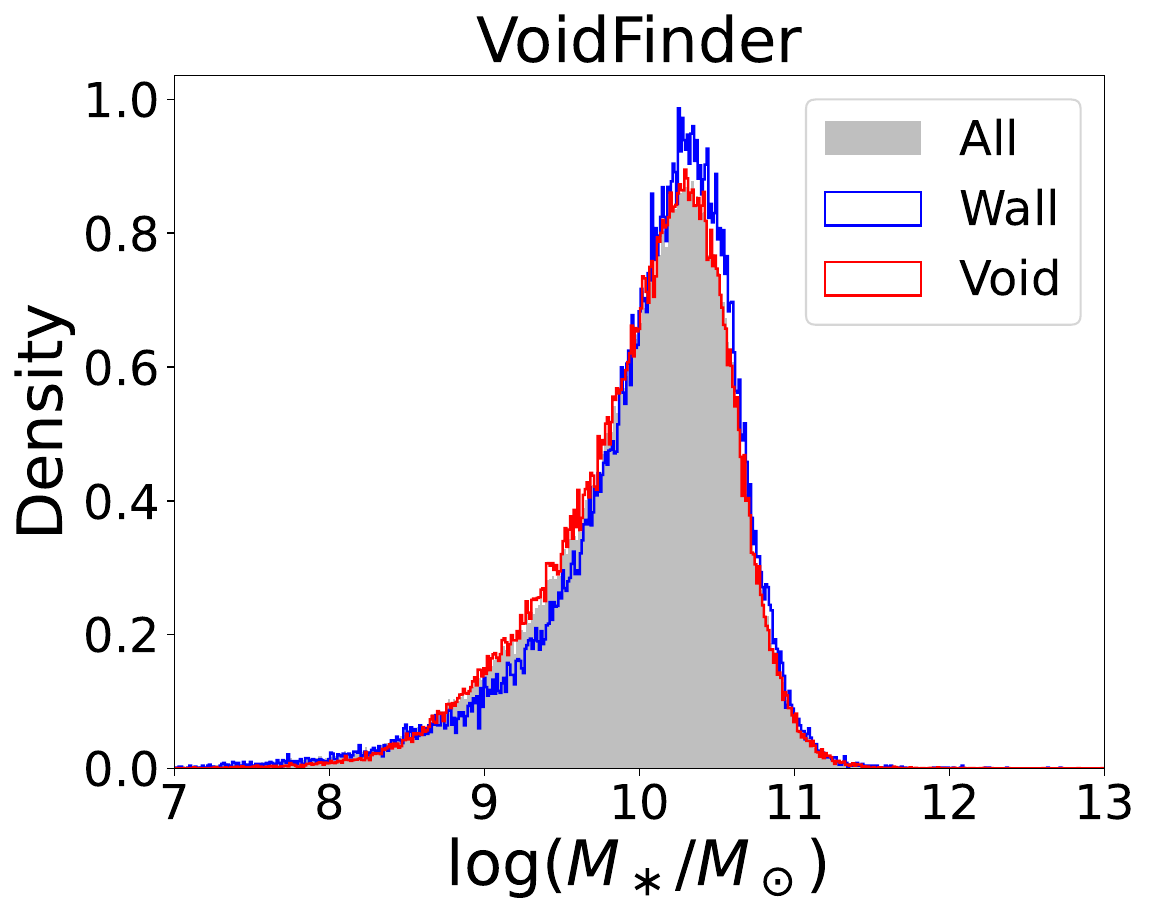}
  \end{subfigure}

  \vspace{0.5cm}

  % Second row
  \begin{subfigure}[t]{0.45\textwidth}
    \includegraphics[width=\linewidth]{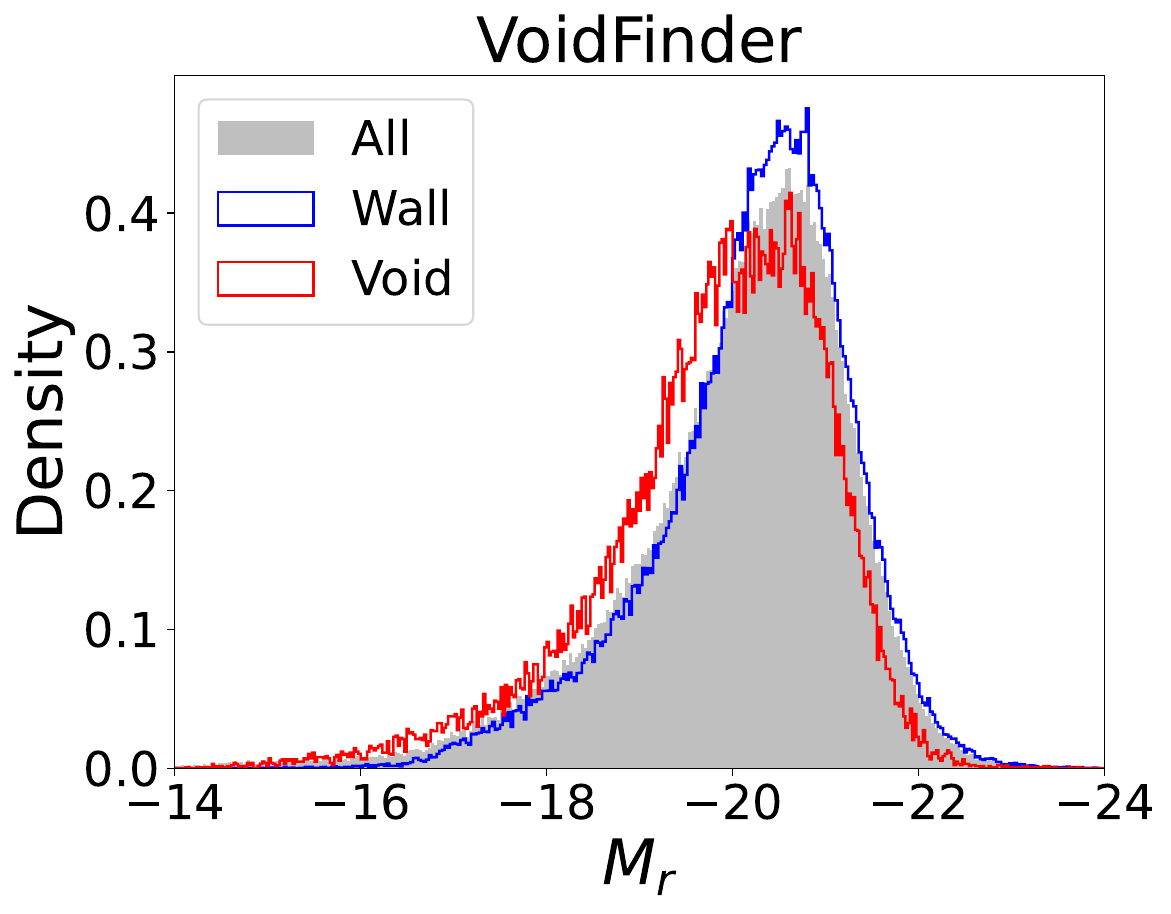}
  \end{subfigure}
  \hfill
  \begin{subfigure}[t]{0.45\textwidth}
    \includegraphics[width=\linewidth]{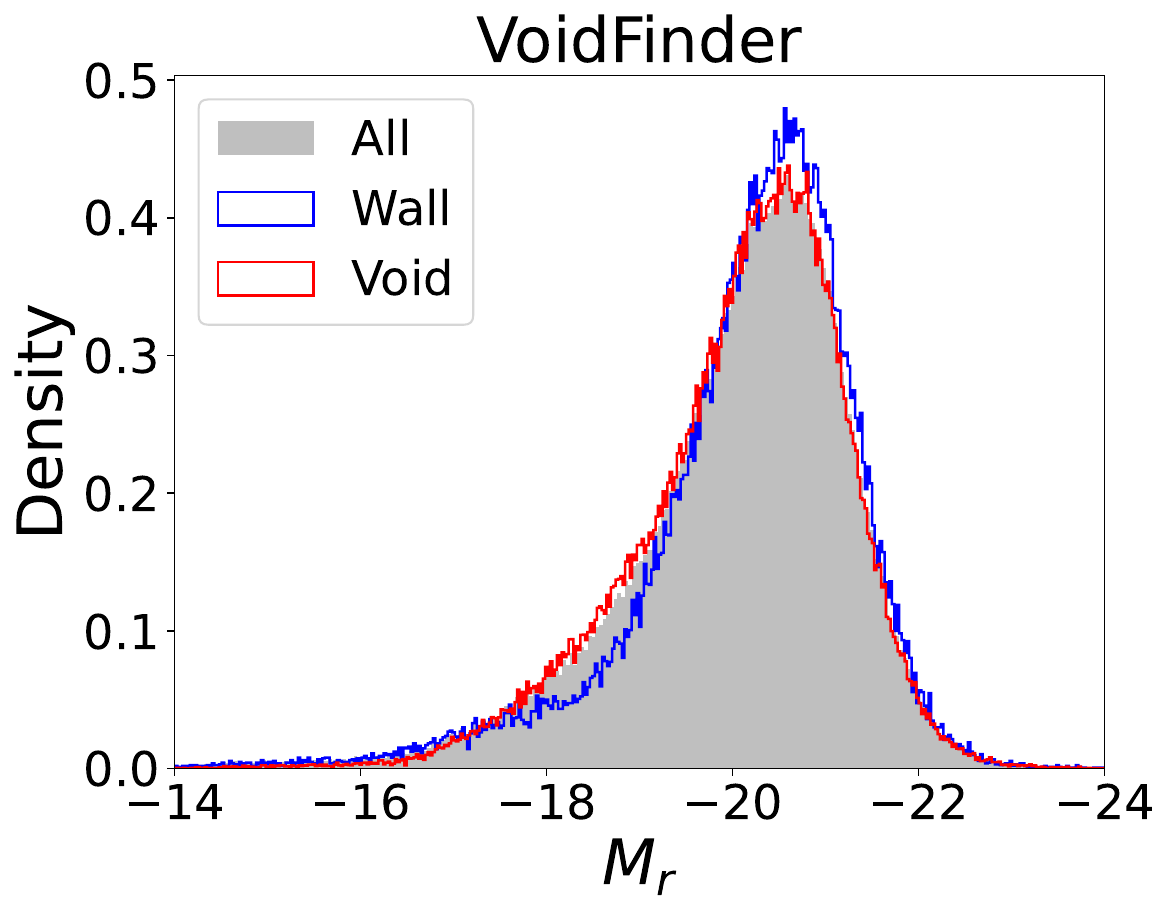}
  \end{subfigure}

  \caption[SDSS DR7 distributions of stellar mass and luminosity]{Stellar mass distribution (top) and luminosity distribution (bottom) of galaxies, separated into void and wall environments according to the VoidFinder (left) and $V_2$ (right) void catalogs. The full galaxy population is shown as a shaded
gray histogram. Blue and red step-line histograms represent wall and void galaxies.}
  \label{fig:sdssmass_mag_grid}
\end{figure}

\begin{figure}%[h]
  \centering
  % First row
  \begin{subfigure}[t]{0.45\textwidth}
    \includegraphics[width=\linewidth]{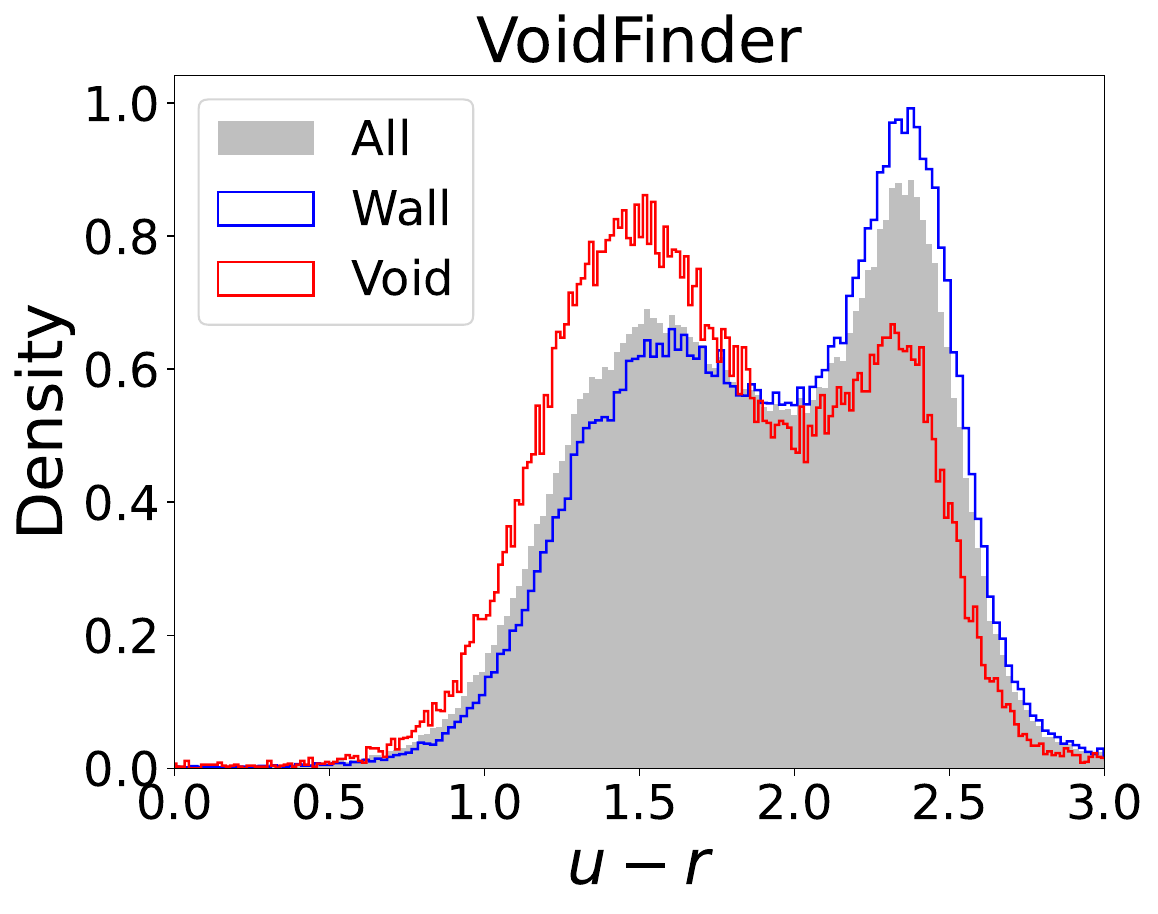}
  \end{subfigure}
  \hfill
  \begin{subfigure}[t]{0.45\textwidth}
    \includegraphics[width=\linewidth]{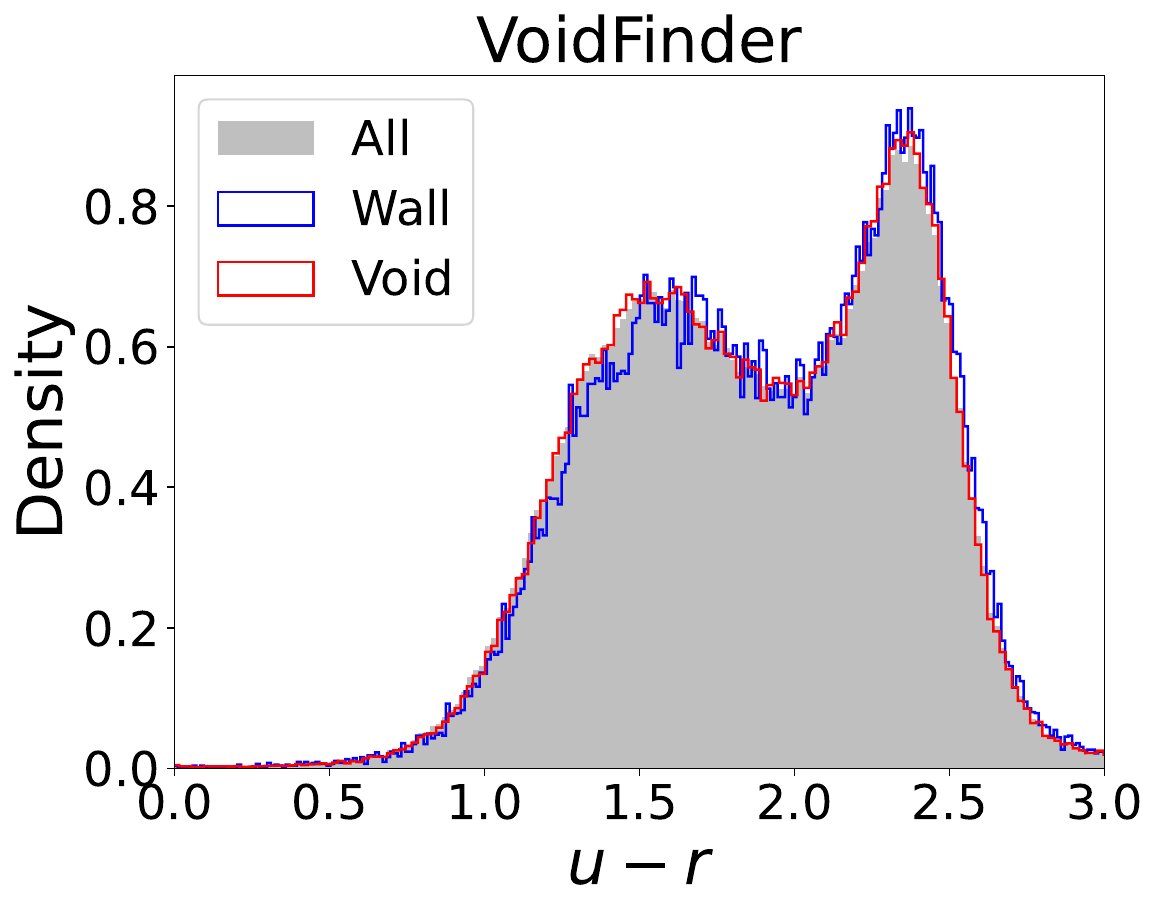}
  \end{subfigure}

  \vspace{0.5cm}

  % Second row
  \begin{subfigure}[t]{0.45\textwidth}
    \includegraphics[width=\linewidth]{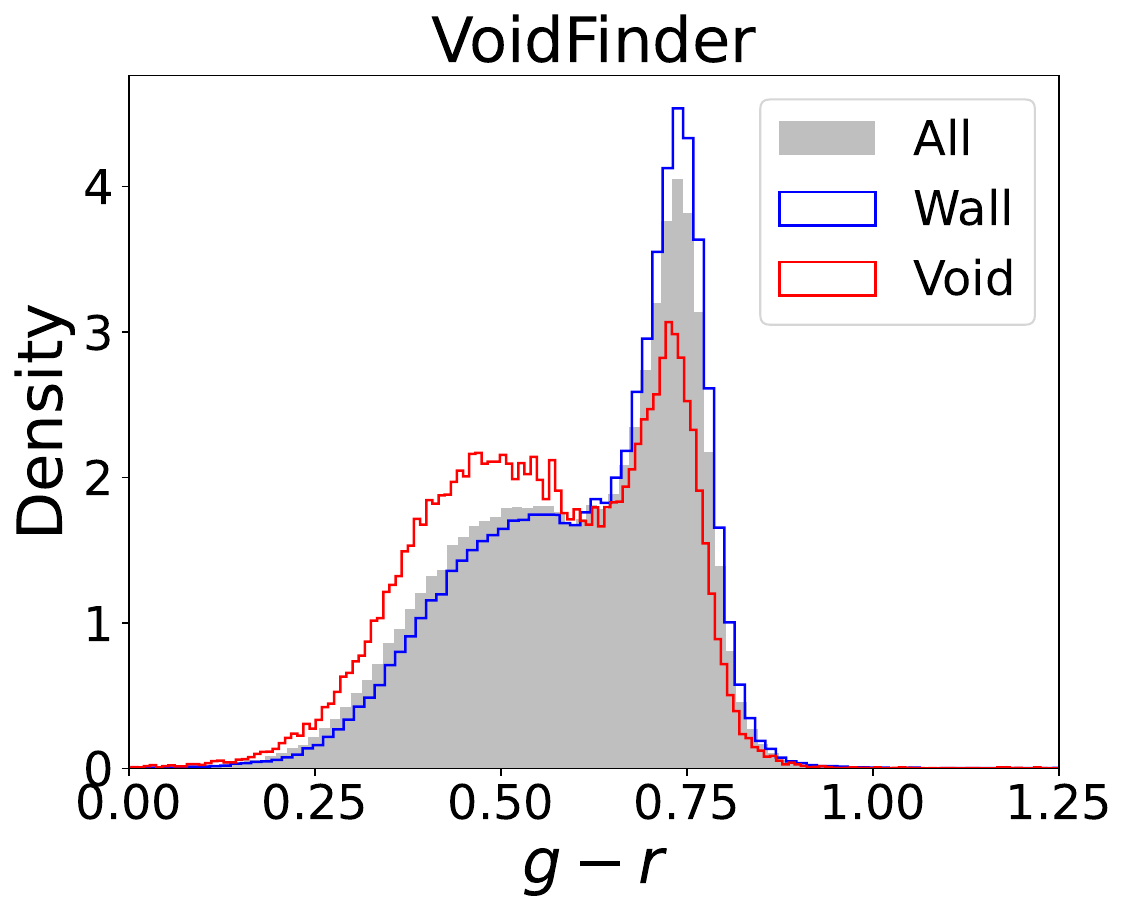}
  \end{subfigure}
  \hfill
  \begin{subfigure}[t]{0.45\textwidth}
    \includegraphics[width=\linewidth]{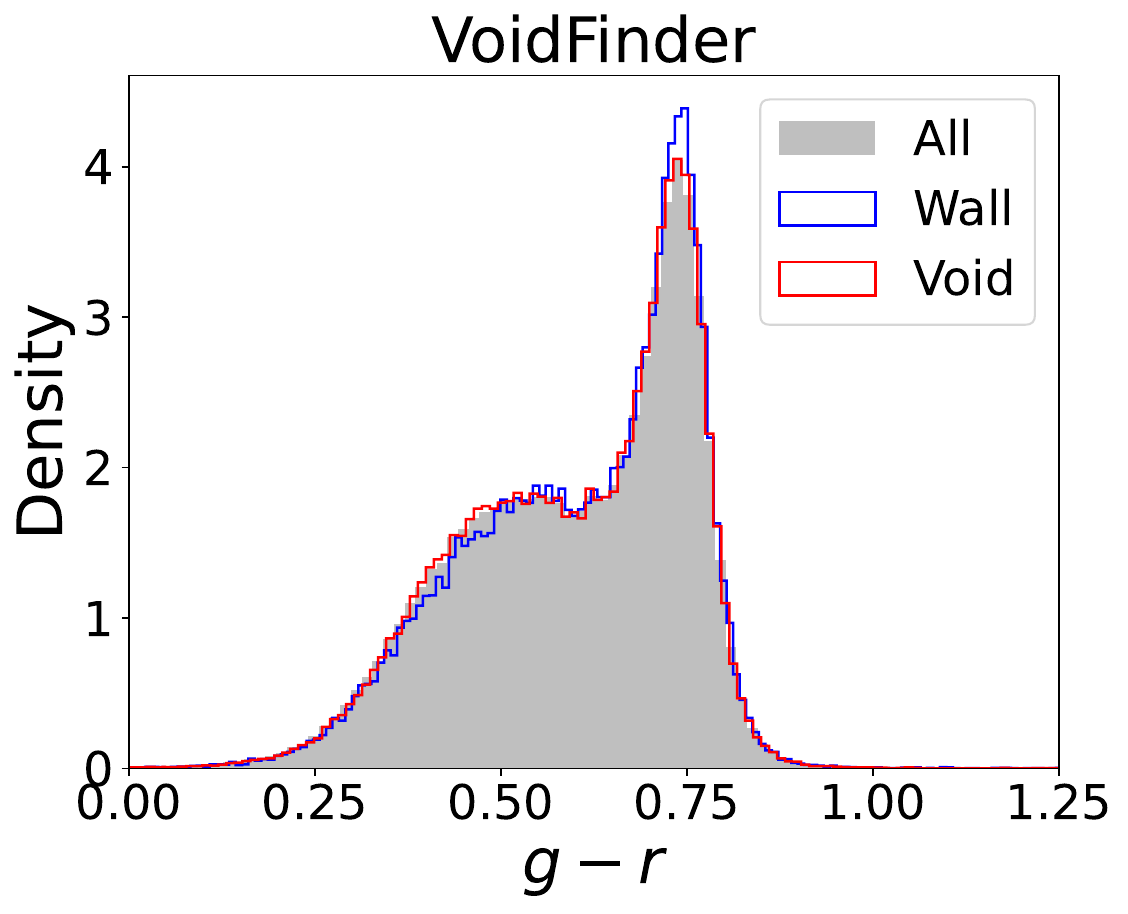}
  \end{subfigure}

  \caption[SDSS DR7 distributions of color]{Color distribution, $u-r$ (top) and $g-r$ (bottom), separated by their environment (void, wall) using
VoidFinder (left column) and $V_2$ (right column). The full galaxy population is shown as a shaded
gray histogram. Blue and red step-line histograms represent wall and void galaxies.}
  \label{fig:sdsscolor_grid}
\end{figure}

\begin{figure}%[h]
  \centering
  % First row
  \begin{subfigure}[t]{0.45\textwidth}
    \includegraphics[width=\linewidth]{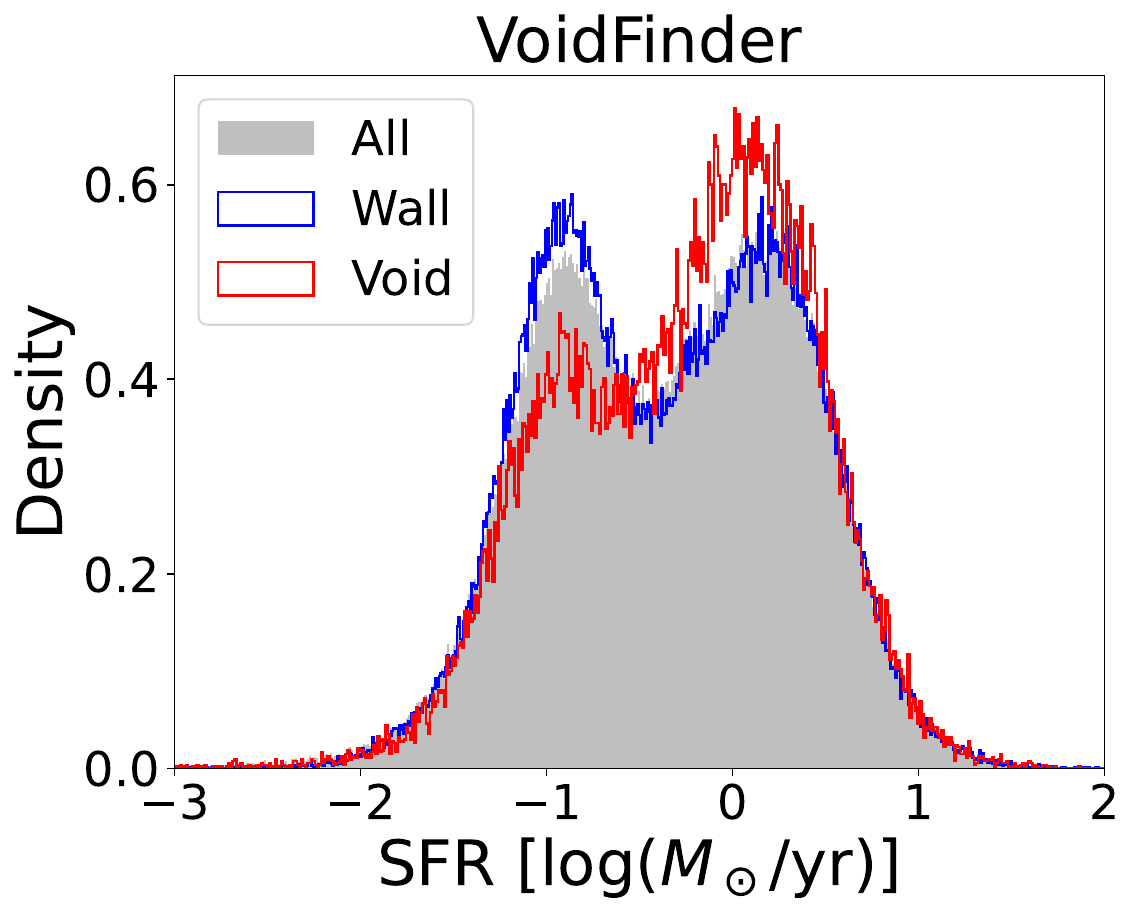}
  \end{subfigure}
  \hfill
  \begin{subfigure}[t]{0.45\textwidth}
    \includegraphics[width=\linewidth]{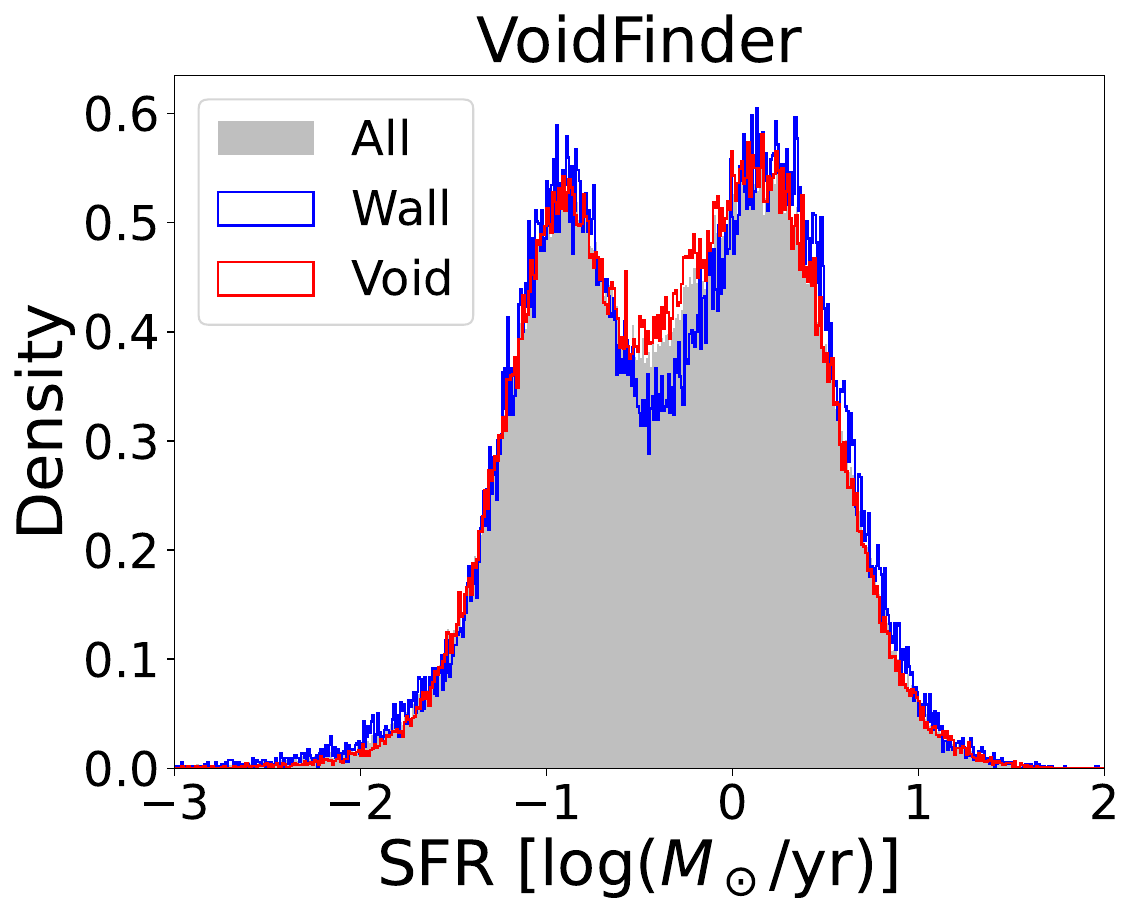}
  \end{subfigure}

  \vspace{0.5cm}

  % Second row
  \begin{subfigure}[t]{0.45\textwidth}
    \includegraphics[width=\linewidth]{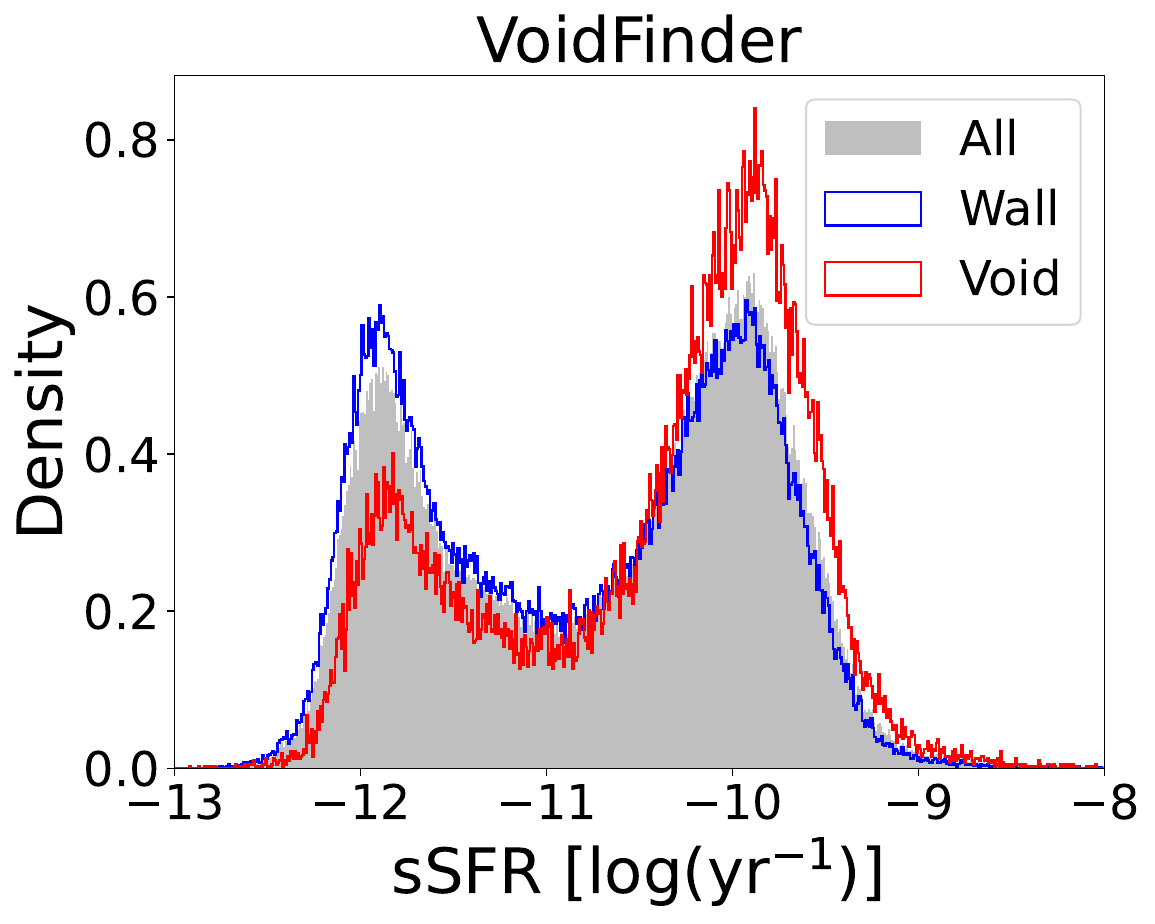}
  \end{subfigure}
  \hfill
  \begin{subfigure}[t]{0.45\textwidth}
    \includegraphics[width=\linewidth]{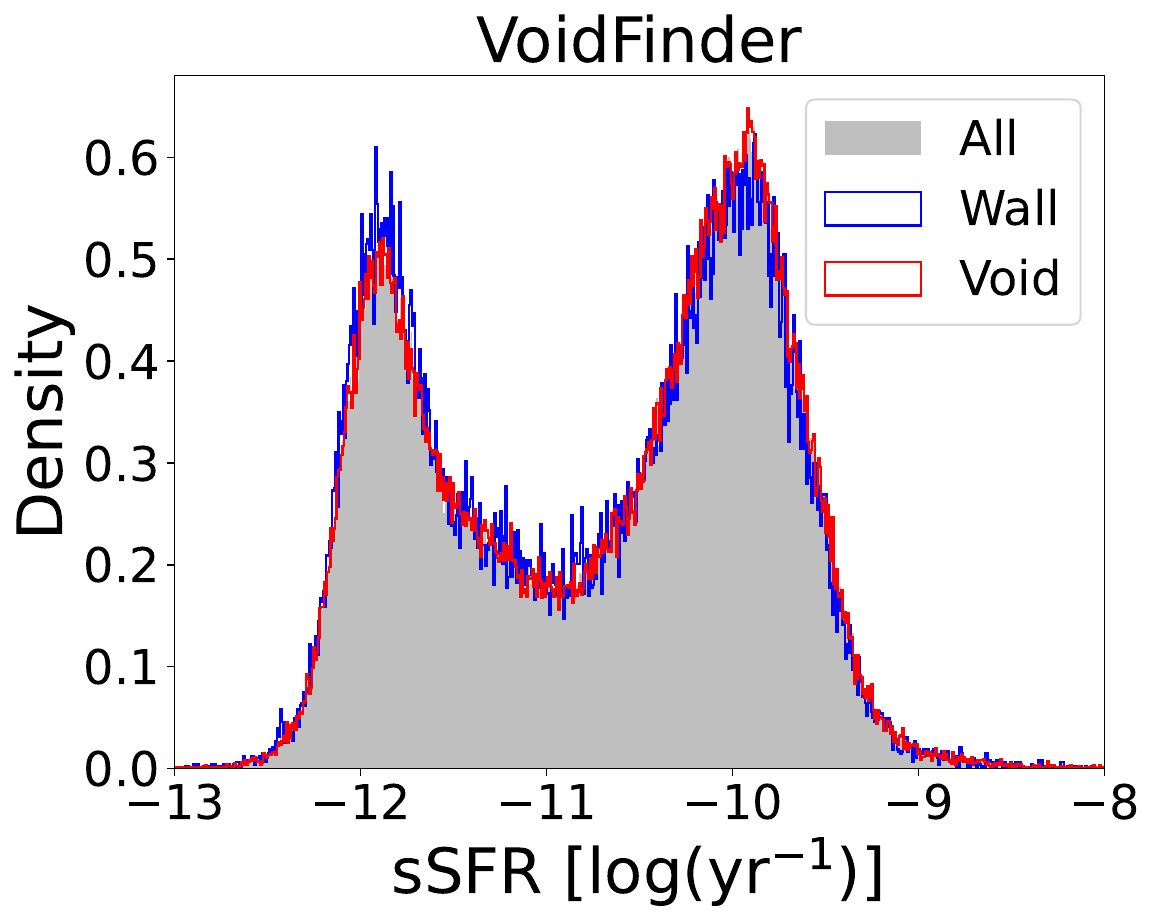}
  \end{subfigure}

  \caption[SDSS DR7 distributions of SFR and sSFR]{Star formation rate (top) and specific star formation rate (bottom), separated by their environment (void, wall) using
VoidFinder (left column) and $V_2$ (right column). The full galaxy population is shown as a shaded
gray histogram. Blue and red step-line histograms represent wall and void galaxies.}
  \label{fig:sdsssfr_grid}
\end{figure}

Figures~\ref{fig:sdssmass_mag_grid}--\ref{fig:sdsssfr_grid} show the distributions of stellar mass, absolute magnitude, $u-r$ and $g-r$ color distributions, SFR, and sSFR, based on the SDSS DR7 data. In each plot, the full galaxy population is shown as a shaded gray histogram. Blue and red step-line histograms represent wall and void galaxies, respectively.

\begin{table}[h]
\centering
\caption{Log Bayes factors for SDSS DR7 under parametric (from \cite{zaidouni2024impact}) and nonparametric Bayesian ($c = 1$, $m = 6$)}
\renewcommand{\arraystretch}{1.2}
\setlength{\tabcolsep}{6pt}
\resizebox{\textwidth}{!}{
\begin{tabular}{llrrrrrr}
\toprule
Method & Algorithm & Stellar Mass & $M_r$ & $g - r$ & $u - r$ & SFR & sSFR \\
\midrule
\multirow{2}{*}{Parametric} 
& VoidFinder      & -1708 & -1248 & -1704 & -1592 & -416 & -1471 \\
& $V^2$ REVOLVER  & -203  & -308  & -58 & -29 & -67 & -19 \\
\midrule
\multirow{2}{*}{Nonparametric} 
& VoidFinder      & -3896 & -2901 & -3847 & -3608 & -895 & -3247 \\
& $V^2$ REVOLVER  & -280  & -547  & 123   & 80    & -58 & 161 \\
\bottomrule
\end{tabular}}
\label{tab:bayes_factors}
\end{table}

Table~\ref{tab:bayes_factors} summarizes the log Bayes factor results under both the parametric (from \cite{zaidouni2024impact}) and nonparametric Bayesian approaches for SDSS DR7 data. Across nearly all galaxy properties, our nonparametric method produces Bayes factors that are more strongly negative under VoidFinder, indicating stronger evidence in favor of the alternative hypothesis $H_1$ (i.e., the distributions differ between void and wall galaxies). For instance, under VoidFinder, the log Bayes factor for stellar mass drops from $-1708$ (parametric) to $-3896$ (nonparametric), amplifying the evidence for distributional differences.

Interestingly, for $V^2$ REVOLVER, the shift is more nuanced. In some cases, the parametric and nonparametric approaches lead to qualitatively different conclusions. For example, under $V^2$ in the case of specific star formation rate (sSFR), the parametric Bayes factor is $-19$, suggesting weak support for $H_1$, whereas the nonparametric Bayes factor is $+161$, indicating positive support for the null hypothesis $H_0$. This discrepancy highlights that while the nonparametric model is generally more sensitive to differences between distributions, it may become conservative when differences between the distributions are subtle.

\section{Comparison between void and wall galaxies through DESI DR1 BGS}

To further evaluate the effectiveness of the nonparametric Bayesian test in identifying distributional differences between void and wall galaxies, we apply the method to a larger dataset: the DESI DR1 BGS. Building upon our findings from SDSS DR7, where the nonparametric approach demonstrated greater sensitivity, we now explore how galaxy properties vary across cosmic environments using DESI data and two widely used void classification algorithms.

Of the 3,378,897 DESI DR1 BGS galaxies that meet our selection criteria, the VoidFinder algorithm classifies 376,788 galaxies as void galaxies and 493,703 as wall galaxies, while the $V^2$ REVOLVER algorithm identifies 812,160 galaxies as void galaxies and 69,378 as wall galaxies.

\begin{figure}%[h]
  \centering
  % First row
  \begin{subfigure}[t]{0.45\textwidth}
    \includegraphics[width=\linewidth]{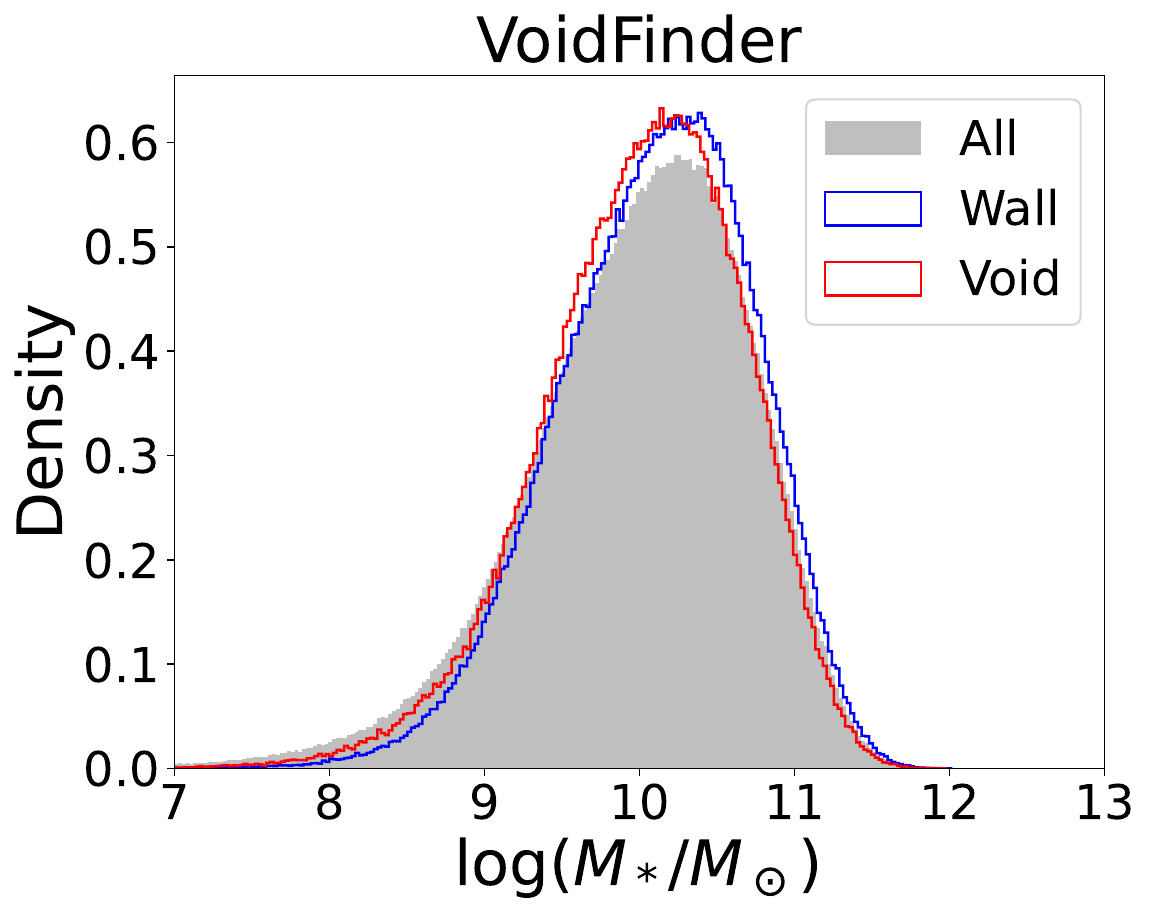}
  \end{subfigure}
  \hfill
  \begin{subfigure}[t]{0.45\textwidth}
    \includegraphics[width=\linewidth]{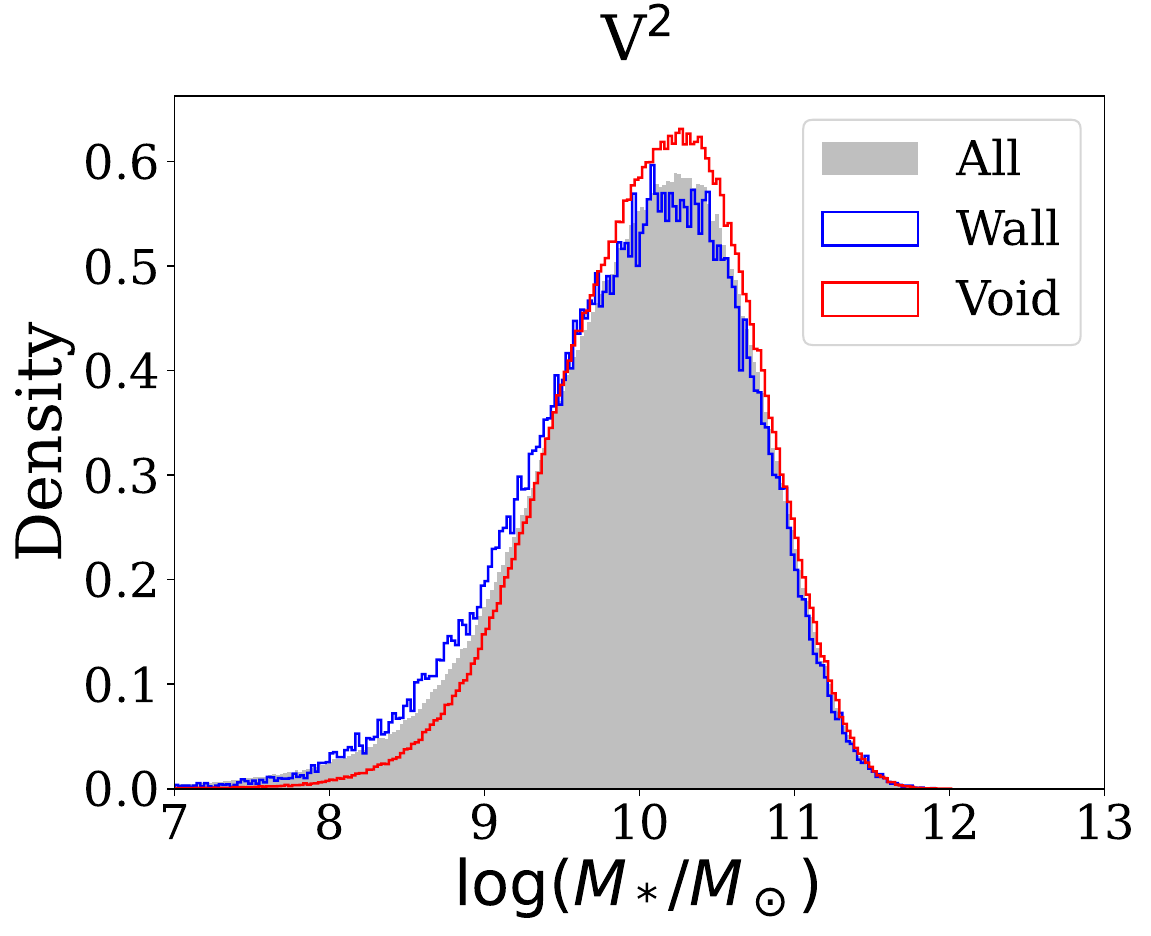}
  \end{subfigure}

  \vspace{0.5cm}

  % Second row
  \begin{subfigure}[t]{0.45\textwidth}
    \includegraphics[width=\linewidth]{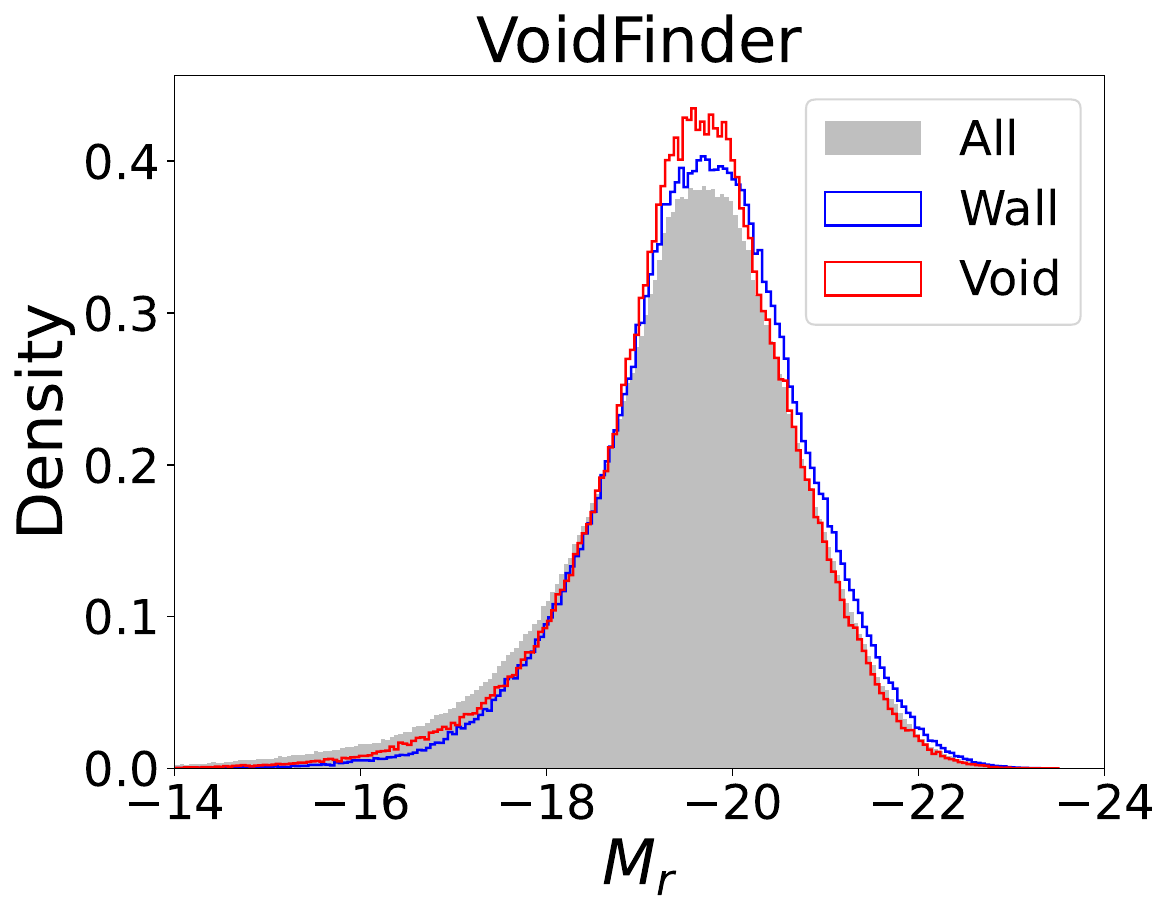}
  \end{subfigure}
  \hfill
  \begin{subfigure}[t]{0.45\textwidth}
    \includegraphics[width=\linewidth]{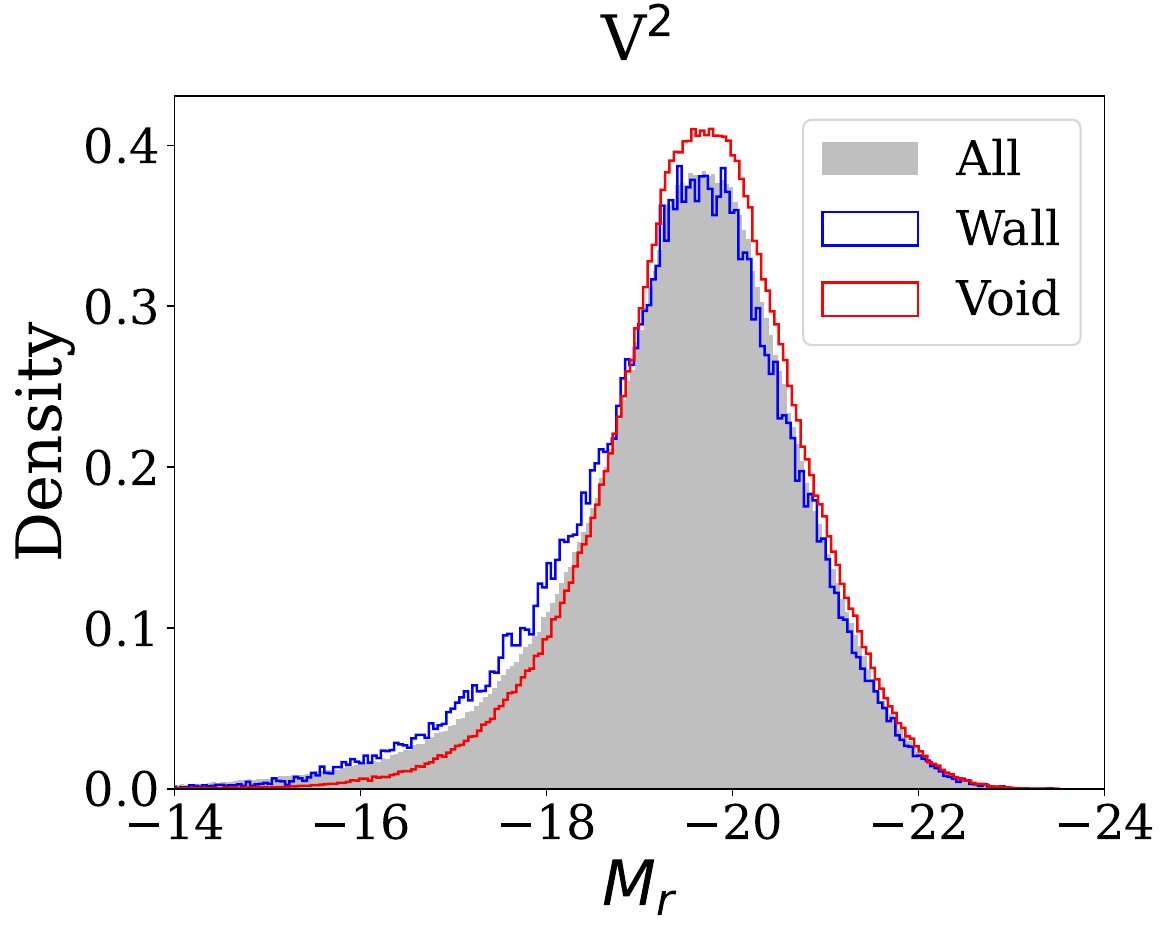}
  \end{subfigure}

  \caption[DESI BGS distributions for stellar mass and luminosity]{Stellar mass distribution (top) and luminosity distribution (bottom) of galaxies, separated into void and wall environments according to the VoidFinder (left) and $V_2$ (right) void catalogs. The full galaxy population is shown as a shaded
gray histogram. Blue and red step-line histograms represent wall and void galaxies.}
  \label{fig:mass_mag_grid}
\end{figure}

\begin{figure}%[h]
  \centering
  % First row
  \begin{subfigure}[t]{0.45\textwidth}
    \includegraphics[width=\linewidth]{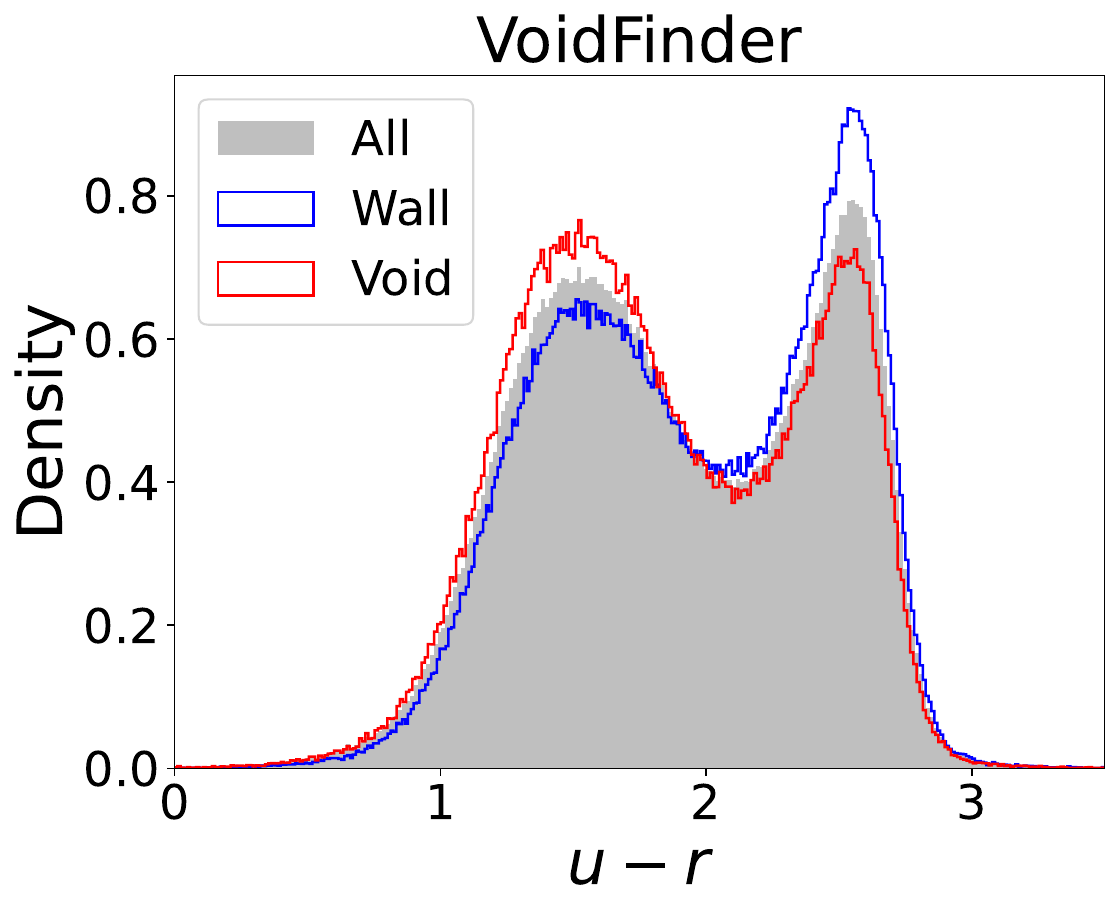}
  \end{subfigure}
  \hfill
  \begin{subfigure}[t]{0.45\textwidth}
    \includegraphics[width=\linewidth]{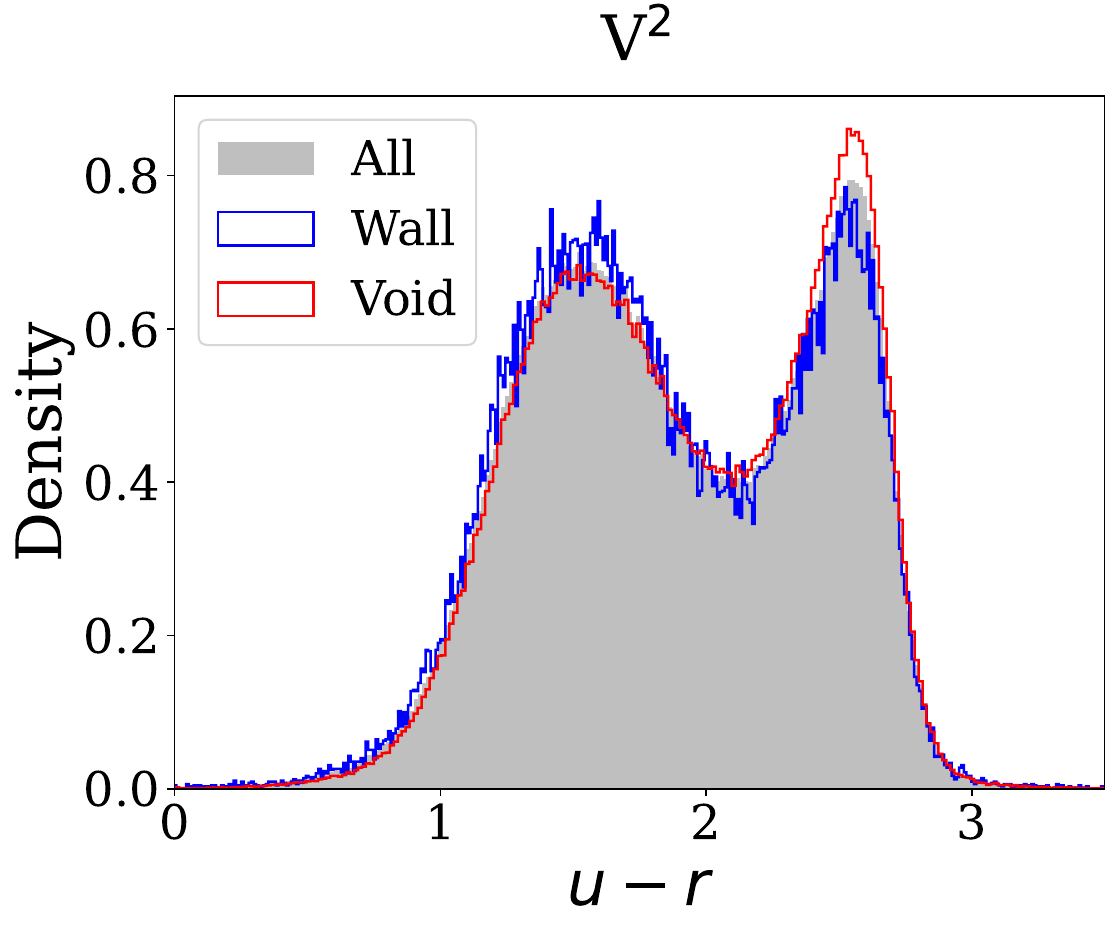}
  \end{subfigure}

  \vspace{0.5cm}

  % Second row
  \begin{subfigure}[t]{0.45\textwidth}
    \includegraphics[width=\linewidth]{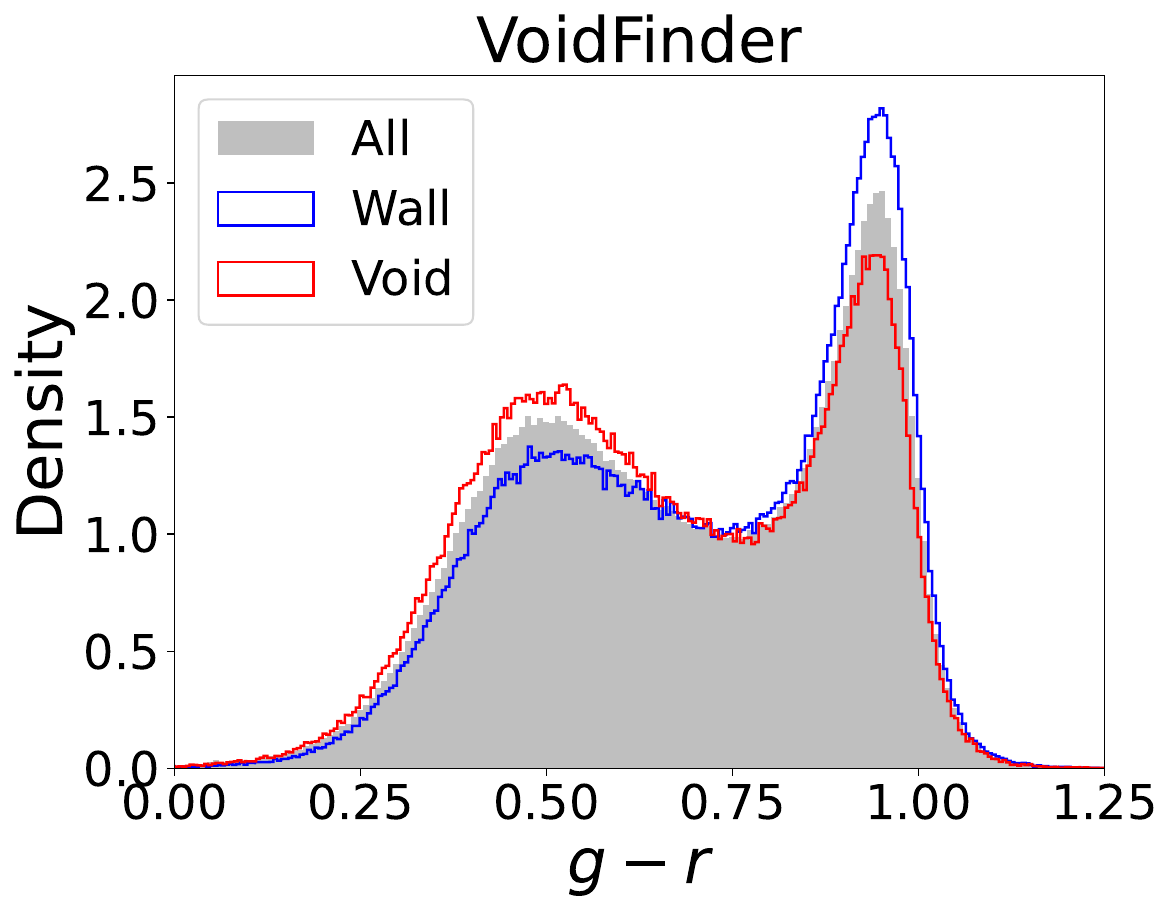}
  \end{subfigure}
  \hfill
  \begin{subfigure}[t]{0.45\textwidth}
    \includegraphics[width=\linewidth]{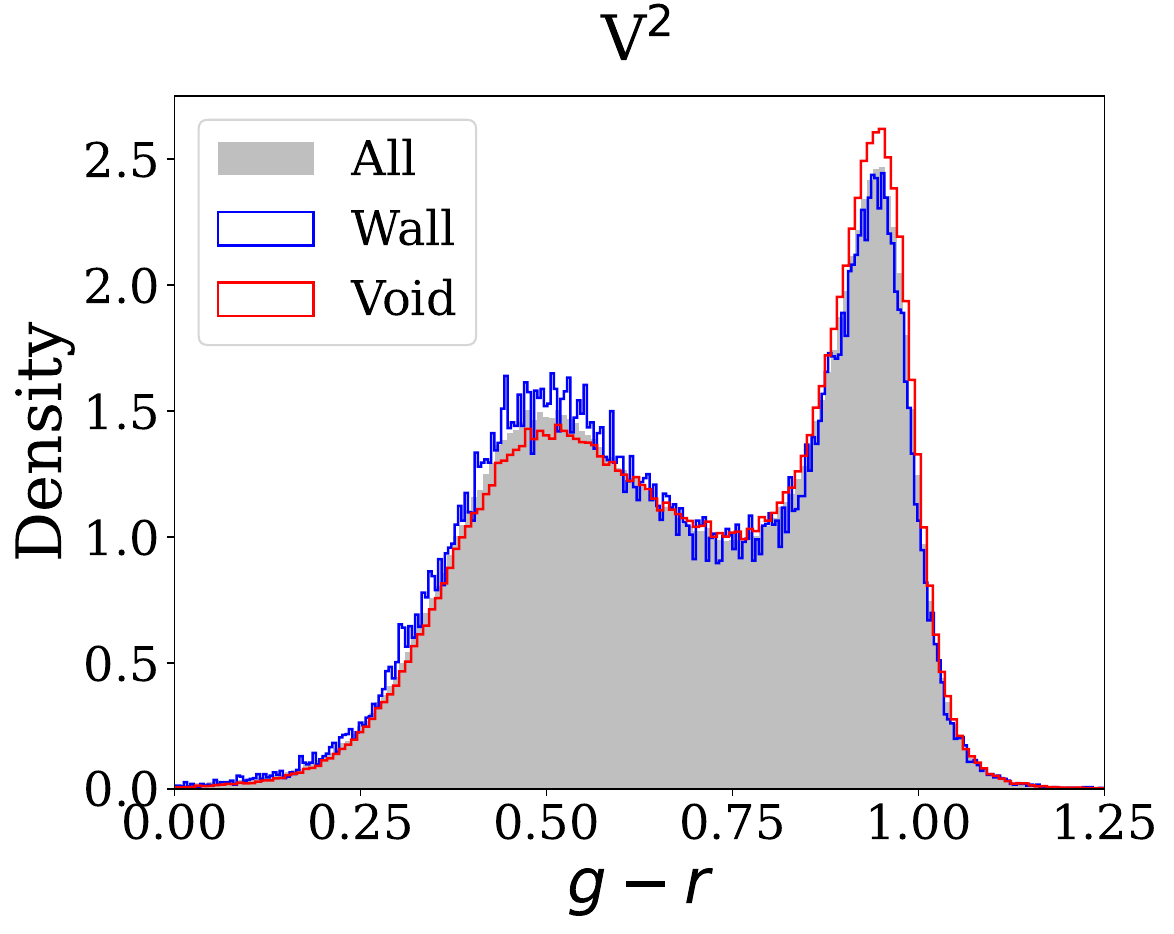}
  \end{subfigure}

  \caption[DESI BGS distributions for color]{Color distribution, $u-r$ (top) and $g-r$ (bottom), separated by their environment (void, wall) using
VoidFinder (left column) and $V_2$ (right column). The full galaxy population is shown as a shaded
gray histogram. Blue and red step-line histograms represent wall and void galaxies.}
  \label{fig:color_grid}
\end{figure}

\begin{figure}%[h]
  \centering
  % Top row (Halpha EW)
  \begin{subfigure}[t]{0.45\textwidth}
    \includegraphics[width=\linewidth]{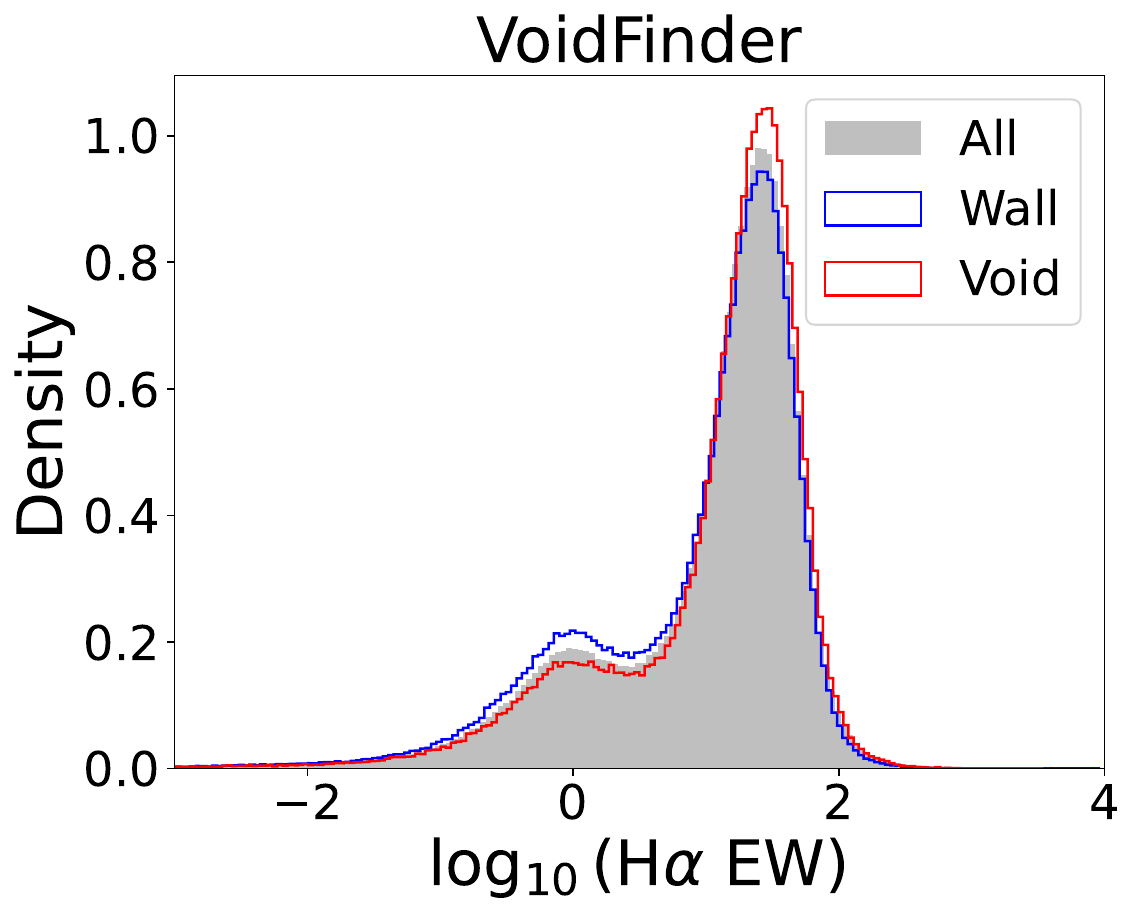}
  \end{subfigure}
  \hfill
  \begin{subfigure}[t]{0.45\textwidth}
    \includegraphics[width=\linewidth]{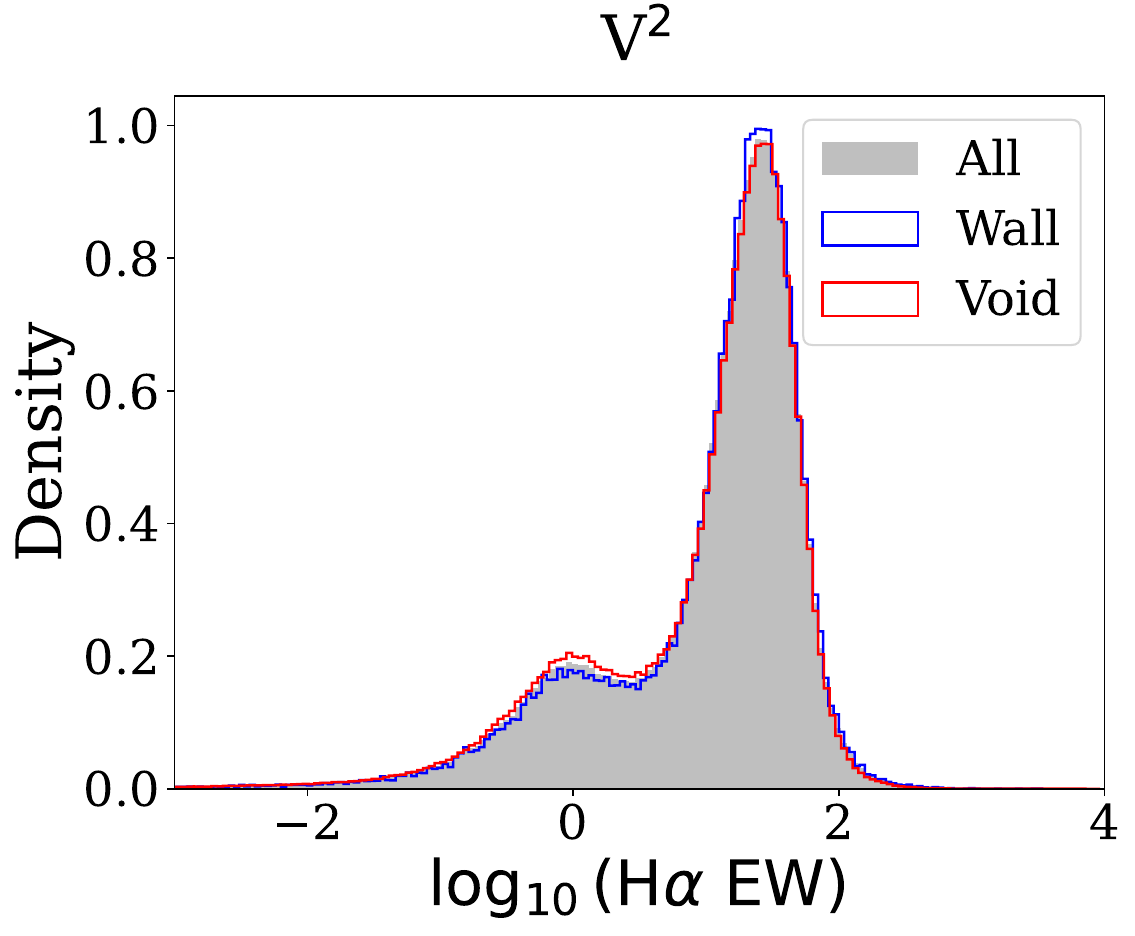}
  \end{subfigure}

  \caption[DESI BGS distributions of H$\alpha$]{Distribution of log H$\alpha$ Equivalent Width according to VoidFinder (left) and $V_2$ (right). The full galaxy population is shown as a shaded
gray histogram. Blue and red step-line histograms represent wall and void galaxies.}
  \label{fig:halpha_grid}
\end{figure}

Figures~\ref{fig:mass_mag_grid}-\ref{fig:halpha_grid} show the distributions of stellar mass, absolute magnitude, $u-r$ and $g-r$ color distributions, and log H$\alpha$ equivalent width as a proxy for log(sSFR), based on the DESI DR1 BGS data.

For distributions generated with the VoidFinder classification, we observe systematic shifts between void and wall galaxy populations across all properties. The stellar mass and absolute magnitude distributions of void galaxies shift leftward compared to wall galaxies, indicating that void galaxies tend to be less massive and fainter. The color distributions in both \( u - r \) and \( g - r \) also shift to the left for void galaxies, reflecting their bluer colors and younger stellar populations. In contrast, the distributions of \(\log(\mathrm{H}\alpha\text{ EW})\), used as a proxy for \(\log(\mathrm{sSFR})\), shift rightward for void galaxies, indicating higher star formation activity. These shifts are consistent with the expectation that galaxies in sparse regions are fainter, less massive, bluer, and more star-forming than galaxies in denser regions.

In contrast, the $V^2$ REVOLVER classification results in more overlapping distributions between void and wall galaxies, particularly in color and \(\log(\mathrm{H}\alpha\text{ EW})\). This observation is reinforced quantitatively by the nonparametric Bayes factors in Table~\ref{tab:DESInonpara_bf}. While VoidFinder shows strong evidence for distributional differences (highly negative log Bayes factors) across all five properties, $V^2$ REVOLVER yields substantially weaker signals, especially for log(H$\alpha$ EW), where its log Bayes factor is closer to zero. 

\begin{table}[h]
\centering
\caption{Log Bayes factor for DESI DR1 BGS under nonparametric Bayesian($c = 1$, $m = 6$)}
\label{tab:DESInonpara_bf}
\begin{tabular}{ccccccc}
\toprule
Algorithm & Stellar Mass & $M_r$ & $g - r$ & $u - r$  & H$\alpha$ EW \\
\midrule
\text{VoidFinder} & -1899 & -1305 & -3946 & -3583 &  -2401 \\
\text{$V^2$ REVOLVER}       & -1404  & -1600  &  -190  &  -224  & -76 \\
\bottomrule
\end{tabular}
\end{table}

These differences lead to an important consequence: galaxies classified as being in VoidFinder voids tend to show stronger statistical and physical differences from wall galaxies, whereas those in $V^2$ REVOLVER voids often resemble wall galaxies in their properties. This outcome aligns with our expectations, as discussed in Chapter~\ref{sec:void-methods}, where we compare the structural definitions of VoidFinder and $V^2$ REVOLVER.

\chapter{Conclusion}

In this study, we explored the influence of the cosmic environment on galaxy evolution by comparing galaxy properties in void and wall regions using two major galaxy redshift surveys: SDSS DR7 and DESI DR1 BGS. A key focus of our study was the application of a nonparametric Bayesian method based on Pólya tree priors, which provides a flexible and robust framework for comparing probability distributions without strong parametric assumptions.

We began with a simulation-based sensitivity analysis of the nonparametric Bayesian test, demonstrating its strong ability to detect differences in both mean and variance between distributions. Additionally, we examined the effect of the prior precision parameter $c$ on test behavior, concluding that $c = 1$ offers a practical balance between responsiveness and robustness for our use cases. 

We then applied this method to SDSS DR7 data, comparing the log Bayes factors with those obtained using a parametric Bayesian test adapted from \cite{zaidouni2024impact}. Under the VoidFinder classification, the nonparametric approach yields consistently more negative Bayes factors across all galaxy properties, indicating stronger and more decisive evidence for distributional differences between void and wall galaxies. In contrast, results under the $V^2$ REVOLVER classification are more nuanced: in several cases, the nonparametric test supports the null hypothesis while the parametric test favors the alternative. These findings underscore the increased sensitivity of nonparametric methods under well-separated distributions, while also highlighting their conservative behavior when differences are subtle.

Next, we applied this method to the DESI DR1 BGS sample. Through histogram comparisons and Bayes factor analysis, we observed that galaxies in VoidFinder-defined voids consistently show fainter magnitudes, lower stellar masses, bluer colors, and higher star formation activity. In contrast, galaxies classified as void galaxies by the $V^2$ REVOLVER algorithm do not differ substantially from wall galaxies. This outcome is consistent with the structural differences between the algorithms, as discussed in Chapter~\ref{sec:void-methods}, where we noted that $V^2$ REVOLVER voids often include wall-like structures within the void volume and combine adjacent dynamically distinct void regions into single voids \cite{rincon2024desivast}.

Overall, our findings affirm that environmental effects on galaxy evolution are detectable and statistically significant, particularly when using the VoidFinder algorithm to define cosmic voids and nonparametric Bayesian tools to compare galaxy populations. 

We plan to further investigate the effect of tree depth \( m \) on the behavior of the nonparametric Bayes factor. While our preliminary results suggest that variations in \( m \) do not drastically change the conclusions of the hypothesis test, we observed that increasing \( m \) can still introduce slight fluctuations in the Bayes factor values, even when all other parameters are held constant. Moreover, deeper trees substantially increase model complexity and computational cost. In future work, we aim to explore how different values of \( m \) influence sensitivity and robustness, and to identify a practical range of \( m \) that balances performance with computational efficiency.

% \appendix

% \include{chapters/99-appendix}

\singlespacing
\printbibliography[heading=bibintoc]

\end{document}